\documentclass[superscriptaddress,showpacs,aps,twocolumn,prb,floatfix]{revtex4}
\usepackage{eurosym}
\usepackage{amssymb}
\usepackage{amsmath}
\usepackage{amsfonts}
\usepackage{amsbsy}
\usepackage{graphicx}
\usepackage{bm}
\usepackage{color}
\usepackage{hyperref}
\usepackage{titlesec}

\begin{document}

\title{Fully compensated Kondo effect for a two-channel spin $S=1$ impurity}

\author{G. G. Blesio}
\affiliation{Instituto de F\'{\i}sica Rosario (CONICET) and Universidad Nacional de Rosario, 
Bv. 27 de Febrero 210 bis, 2000 Rosario, Argentina}

\author{L. O. Manuel}
\affiliation{Instituto de F\'{\i}sica Rosario (CONICET) and Universidad Nacional de Rosario, 
Bv. 27 de Febrero 210 bis, 2000 Rosario, Argentina}

\author{A. A. Aligia}
\affiliation{Centro At\'{o}mico Bariloche and Instituto Balseiro, Comisi\'{o}n Nacional
de Energ\'{\i}a At\'{o}mica, CONICET, 8400 Bariloche, Argentina}

\author{P. Roura-Bas}
\affiliation{Centro At\'{o}mico Bariloche and Instituto Balseiro, Comisi\'{o}n Nacional
de Energ\'{\i}a At\'{o}mica, CONICET, 8400 Bariloche, Argentina}

\begin{abstract}
We study the low-temperature properties of the generalized Anderson impurity model in which two localized
configurations, one with two doublets and the other with a triplet, are mixed by two degenerate conduction channels.
By using the numerical renormalization group and the non-crossing approximation, we analyze the impurity entropy, 
its spectral density, and the equilibrium conductance for several values of the model parameters. Marked differences 
with respect to the conventional one-channel spin $s=1/2$ Anderson model, that can be traced as hallmarks of an 
impurity spin $S=1$, are found in the Kondo temperature, the width and position of the charge transfer peak, and 
the temperature dependence of the equilibrium conductance. Furthermore, we analyze the rich effects of a single-ion 
magnetic anisotropy $D$ on the Kondo behavior. 
In particular, as shown before, for large enough positive $D$ the system behaves as a ``non-Landau'' Fermi liquid that cannot
be adiabatically connected to a non-interacting system turning off the interactions. For negative $D$ the Kondo effect
is strongly suppressed.
While the model is suitable for the description of a single Ni impurity embedded into an O doped Au chain,
it is a generic one for $S=1$ and two channels and might be realized in other nanoscopic systems. 
\end{abstract}

\pacs{73.23.-b, 71.10.Hf, 75.20.Hr}
 
\maketitle

\section{Introduction} \label{sec_introduction}
The Kondo effect, early found in metals containing magnetic impurities~\cite{hewson97,kondo64}, 
is also frequently observed in low dimensional systems. For instance, transport measurements through 
semiconducting~\cite{goldhaber98,cronenwett98,goldhaber98b,wiel00,grobis08,kretinin11,amasha13} and 
molecular~\cite{liang02,yu05,leuenberger06,oso1,parks07,roch08,oso2,scott09,parks10,florens11,vincent12} 
quantum dots (QDs), at low enough temperatures, exhibit the Kondo phenomena. 
Here, the QD acts as a single magnetic impurity while the contacts play the role of metallic hosts.

The seminal work by Nozi\`{e}res and Blandin~\cite{nozieres80} pointed out the crucial role that the impurity 
and the conduction host orbital structures play in the Kondo physics. 
Therefore, real systems are expected to be modeled 
by Kondo Hamiltonians where an arbitrary spin $S$ is screened by $n$ conducting channels 
(bands with different symmetry) of spin $s=1/2,$ 
and the nature of the ground state depends on the relation between $S$ and $n$.
For $n = 2S$ the models have Fermi liquid ground states, while for $n > 2S$ 
they correspond to non-Fermi-liquids and for $n < 2S$ the systems are singular Fermi liquids~\cite{mehta05}. 
The existence of non-Fermi-liquid ground states requires $SU(n)$ symmetry in the conducting channels, 
which is difficult to achieve in real systems. For instance, in the case of the $S = 1/2$ two-channel model, 
the effect of  symmetry-breaking perturbations was discussed by Sela {\it et al.}~\cite{sela11}
It is also found that the presence of magnetic anisotropy can drastically modify the low-temperature 
properties~\cite{cornaglia11,dinapoli13,dinapoli14,zitko08}.

The multiorbital Kondo physics can be found in molecular QDs, which have rich inner electronic structures 
with magnetic orbitals coupled in a such a way that the resulting spin is sometimes larger than the usual $s=1/2$. 
For instance, the underscreened Kondo effect, corresponding to the case $n < 2S$, has been experimentally and 
theoretically investigated for molecules with spin 
$S = 1$~\cite{parks07,roch08,florens11,cornaglia11,roura09,logan09,roura10,barral17},
and larger spin~\cite{barral17,jacob13,expmn}. In other systems, the spin ~\cite{oso1,orma,teodspin} 
and also the anisotropy 
$D$~\cite{teodspin,teodneg} can be manipulated.
On the other hand, a Co impurity in an O-doped Au chain behaves as a QD with 
$S=3/2$~\cite{dinapoli13,dinapoli13b,dinapoli14}. In this system, two conducting gold channels ($5d_{xz},$ $5d_{yz}$), 
degenerate by symmetry, screen only two electrons of the Co atom ($3d_{xz},$ $3d_{yz}$), 
while the electron on the $3d_{xy}$ orbital is unaffected by the Au bands. 
In the presence of a single-ion magnetic anisotropy $DS_z^2,$ with $D > 0$, the effective impurity spin becomes $S=1/2$ and
the corresponding scenario was found to be the overscreened Kondo effect, $n>2S ~ 
(n=2,\;S=1/2)$~\cite{dinapoli13,dinapoli13b,dinapoli14}.

\begin{figure}[ht]
\begin{center}
\includegraphics*[width=\columnwidth]{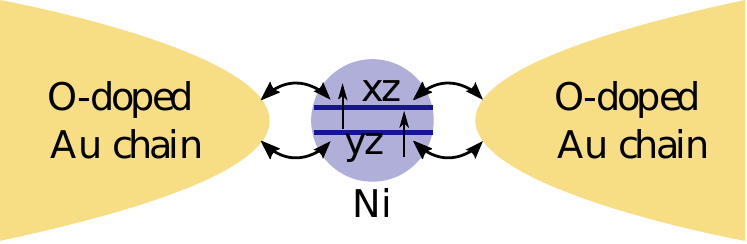}
\caption{Scheme of the system proposed in Ref. \onlinecite{dinapoli15}.
The Ni ion has two holes, with total spin $S=1$, that can jump to conduction bands of the same symmetry 
either to the left or to the right.}
\label{scheme}
\end{center}
\end{figure}

A difference between the bulk systems and the low dimensional QD ones is that while in the former 
the fully screened scenario $n = 2S$ is frequent~\cite{nevidomskyy09}, it seems rare in the latter.
However, as stated above, there are several studies on nanoscopic systems with $S>1/2$ and also with degenerate 
orbitals~\cite{dinapoli13,dinapoli13b,mina,joaq,moro} so that this scenario is expected 
to appear in the future.
In particular, recently a realization of the fully compensated high spin Kondo phenomena 
in a low-dimensional system has been proposed~\cite{dinapoli15}. 
It was shown that a Ni impurity, within an O doped Au chain, has two holes in the degenerate 
$3d_{xz,yz}$ orbitals, coupled to $S = 1$ because of a large Hund's interaction~\cite{dinapoli15,barral17}. 
The coupling between the $5d_{xz,yz}$ Au bands (which doped with O cross the Fermi level) 
and the $3d_{xz,yz}$ of Ni states leads to a two-channel $S = 1$ Kondo effect. 
A scheme of the system is presented in Fig. \ref{scheme}.
Based on first-principles calculations, the electronic structure was studied and 
effective Anderson- and Kondo-like Hamiltonians were derived in Ref. \onlinecite{dinapoli15}. 
Using model parameters, an experimentally accessible Kondo temperature $T_K \sim 70$ K 
was estimated (see Supplemental Material of Ref.~\cite{blesio18}). 
More recently we have shown that the model has a topological phase transition as a function
of the impurity single-ion magnetic anisotropy $D$~\cite{blesio18}.
However, so far, the general properties of the model have not been studied.

While the Kondo screening of a spin $S=1/2$ has been the focus of an intense research since the 70s, 
both experimentally and theoretically, this is not the case of the full screening of the $S = 1$ case. 
For instance, the long-known ``Kondo resonance narrowing problem'',  that is, the exponentially decrease of the 
Kondo scale with the impurity spin $S$~\cite{daybell68}, has not received much attention until the recent work by 
Nevidomskyy and Coleman~\cite{nevidomskyy09}.

In this work, in order to deepen the understanding of the high-spin fully compensated Kondo physics and to 
serve as a guide to experimental characterizations, we present calculations of the model presented in Ref. \onlinecite{dinapoli15} for valence fluctuations 
between a configuration with one particle (two doublets) and another one with two particles (a triplet). This model 
corresponds to an impurity spin $S=1$ screened by two degenerate conduction channels. 
The particles can be electrons or holes. To solve the model, we use two complementary methods:
the numerical renormalization-group (NRG) and the non-crossing approximation (NCA).  

For $D=0$, we present results for the impurity entropy as a function of 
temperature $S(T)$, the impurity occupancy as a function of the impurity level $\epsilon_d$, 
the impurity spectral density $\rho(\omega)$ 
for different $\epsilon_d$ including a study of the position and width of the charge-transfer peak near $\epsilon_d$, 
the Kondo temperature as a function of $\epsilon_d$, and the conductance $S(T)$ as a function of temperature for several
values of $\epsilon_d$. We also show how these results are modified by anisotropy $D$. In Ref. \onlinecite{blesio18} 
we have shown some results for $S(T)$, $\rho(\omega)$ and $G(T)$ for fixed $\epsilon_d$ and several $D \geq 0$, to show that a sharp
jump in these properties take place at a given critical anisotropy $D_c$ due to a topological quantum phase transition.
Here, we add new calculations in the Kondo limit particularly for $D$ near $D_c$ which illustrate 
the sharpness of the transition. 
We also study the case $D < 0$. While for a Ni impurity 
in a Au chain a positive $D$ or the order of a few meV has been calculated~\cite{dinapoli15}, 
one might expect that experimental realizations of the fully screened Kondo model
with $S>1/2$ and $D<0$ appear in the future. 
In particular negative tunable $D$ has been calculated in some systems containing 
phtalocyanine molecules~\cite{teodspin,teodneg}.
The Haldane system Y$_{2}$BaNiO$_{5}$ has negative $D$ \cite{payen} 
and one might expect the same for a Ni impurity in a similar environment.

For $D=0$, we have found, as expected~\cite{nozieres80}, several features that correspond to a Fermi liquid behavior at low 
enough temperatures, for example, the impurity entropy $S_{imp}(T) \rightarrow 0$ as $T \rightarrow 0$.
On the other hand, the impurity spectral density $\rho_{\alpha\sigma}(\omega)$ exhibits a single charge-transfer 
tunneling resonance (the impurity configuration with three particles is absent in the Hamiltonian), and also the 
Kondo one within the Kondo regime. The features of the resonances are discussed in comparison with the corresponding 
one-channel $S=1/2$ case. In particular, we have found that the temperature dependence of the conductance can be used as 
a hallmark to characterize a fully screened $S=1$ impurity. On the other hand, the presence of a single-ion magnetic 
anisotropy $D$ profoundly affects the Kondo physics. As we have shown recently for particular 
parameters~\cite{blesio18}, for positive $D$ there is a topological transition at a critical value $D_c$ 
to another Fermi liquid phase which cannot be adiabatically connected to a non-interacting system. 
Instead for $D<0$, The Kondo physics is preserved but the Kondo energy scale is strongly reduced because 
the remaining two degenerate states of the localized triplet, with projections $\pm 1$ are mixed by an effective spin 
flip of fourth order in the hybridization between localized and conduction states. 

The paper is organized as follows. In section \ref{sec_model} we introduce the model Hamiltonian as well as the NRG and 
NCA approaches.  In section \ref{sec_numerical} the numerical solution of the model is presented, for several values of the
model parameters including the particular case of the Ni-Au-O system considered in Ref. \onlinecite{dinapoli15}.
Finally, in section \ref{sec_conclusions} the conclusions are drawn.

\section{Model Hamiltonian and Methods} \label{sec_model}
The Hamiltonian that describes the system of a magnetic Ni atom in a substitutional position within an Au chain doped 
with a small amount of oxygen can be written by using hole operators, $h_{\sigma}$, related with the electron operators, 
$d_{\sigma}^\dagger$, by $h_{\uparrow}=d_{\downarrow}^\dagger$ and $h_{\downarrow}=-d_{\uparrow}^\dagger$.
This transformation preserves the form of the Hamiltonian and the spin operators, which have the same form 
in both representations. Using these operators for all atoms in the system, 
and retaining only the ground state of the $d^8$ and $d^9$ configurations, 
the model can be written in the form~\cite{dinapoli15}
\begin{eqnarray} \label{ham}
& H & =\sum_{M_{2}}(E_{2}+ D M_{2}^{2})|1M_{2}\rangle \langle 1M_{2}|+
\sum_{\alpha M_{1}}E_1|\alpha M_{1}\rangle \langle \alpha M_{1}| \nonumber\\
&  & + \sum_{\nu k\alpha \sigma }\epsilon _{\nu k}c_{\nu k\alpha \sigma
}^{\dagger }c_{\nu k\alpha \sigma }+ \\
& & \!+\!\!\sum_{M_{1}M_{2}\nu k \alpha \sigma }\!\!V_{\nu\alpha }
\!\langle 1 M_{2}  |\frac{1}{2}\frac{1}{2}M_{1}\sigma \rangle\!
\left( | M_{2}\rangle \langle \alpha M_{1}|c_{\nu k\alpha \sigma }\!+\!\mathrm{H.c.}\! \right),  \nonumber 
\end{eqnarray}
where $E_i$ and  $M_i$ indicate the energies and the spin projections along the chain (chosen as the 
quantization axis) of states with $i=1, 2$ holes in the $3d$ shell of the Ni impurity. 

The ground state configuration of the Ni atom, described by the first term in Eq. (\ref{ham}) 
is found to have two holes in the degenerate $3d_{xz,yz}$ 
orbitals coupled to spin $S=1$ by means of a strong Hund's coupling.
The state with maximum spin projection can be represented by $\vert 1 1 \rangle = 
{h}_{xz\uparrow}^{\dagger}{h}_{yz\uparrow}^{\dagger} \vert 0 \rangle$, where $\vert 0 \rangle$ stands for the 
full $3d^{10}$ configuration and the operator ${h}_{\alpha\sigma}^{\dagger}$ creates a hole with symmetry 
$\alpha=xz, yz$ and spin projection $\sigma$. 
The other relevant states of the Ni $d^8$ configurations can be obtained by using the spin lowering operator on this state.
$D$ represents the single-ion magnetic anisotropy.

The second term contains the relevant states of the $d^9$ configuration with 
$|xz M_{1}\rangle={h}_{yz M_{1}}^{\dagger} \vert 0 \rangle$
and $|yz M_{1}\rangle={h}_{xz M_{1}}^{\dagger} \vert 0 \rangle$.

The third term represents the four conduction bands for the two channels $\alpha$ and conduction leads $\nu$.
The operator $c_{\nu k\alpha \sigma}^{\dagger }$ creates a hole in the a Au-O band 
with symmetry $\alpha$, where $\nu=L, R$ denotes the left or the right side of the Ni atom, respectively.

The last term characterizes the tunneling between the Ni and Au states for each channel. 
The factor $V_{\nu\alpha }$ defines the hybridization for each channel and lead,
and the Clebsch-Gordan coefficients $\langle 1 M_{2}  |\frac{1}{2}\frac{1}{2}M_{1}\sigma \rangle$ 
determine the ratio among different angular momentum projections. 
For the calculations presented here, we take $V_{\nu\alpha }=V$ independent of lead and channel

Note that the presence of O atoms within the conducting chain is a necessary ingredient for the applicability of 
the model to the system, because electronegative O atoms deplete the Au bands and are responsible for the presence 
of $5d_{xz,yz}$ Au bands at the Fermi level~\cite{dinapoli13,dinapoli13b,dinapoli14}. With this condition, 
the tunneling 
mechanism responsible for charge transfer and spin-flip between Ni and Au neighbors is warranted. 

\subsection{\textit{Numerical Renormalization Group}}\label{sec_nrg}
In order to obtain quantitatively reliable results at low energies, we solve our Hamiltonian by means of NRG, 
which is a 
numerically exact technique. In order to use the NRG LJUBLJANA open source code \cite{nrg_code,zitko09}, 
we note that the Hamiltonian in Eq.~(\ref{ham}) can be derived as a particular case of a more general, pure 
fermionic, two-orbital Anderson model given by
\begin{eqnarray} \label{2orb-ham}
\tilde{H}&=& \tilde{H}_{imp} + \tilde{H}_{c} + \tilde{H}_{mix}
\end{eqnarray}
where the impurity Hamiltonian including Coulomb repulsion \cite{aligia13} reads as follows
\begin{eqnarray} \label{imp-ham}
\tilde{H}_{imp}&=&\sum_{\alpha}\epsilon_{\alpha}n_{\alpha} + U\sum_{\alpha}n_{\alpha\uparrow}n_{\alpha\downarrow} +
U'n_{xz}n_{yz}+ \nonumber\\
&&  +J_H \sum_{\sigma\sigma'}h^{\dagger}_{xz\sigma}h^{\dagger}_{yz\sigma'}h_{xz\sigma'}h_{yz\sigma}+ DS_{z}^{2}+\\
&&  +J_H ( h^{\dagger}_{xz\uparrow}h^{\dagger}_{xz\downarrow}h_{yz\downarrow}h_{yz\uparrow}+ \mathrm{H.c.}) \nonumber\\ 
&&\nonumber\\
&=&\sum_{\alpha}\epsilon_{\alpha}n_{\alpha} + U\sum_{\alpha}n_{\alpha\uparrow}n_{\alpha\downarrow} +
(U'-J_H / 2 )n_{xz}n_{yz} +\nonumber\\
&& -2J_H~\vec{S}_{xz} \cdot \vec{S}_{yz} + DS_{z}^{2}+\\
&&  +J_H ( h^{\dagger}_{xz\uparrow}h^{\dagger}_{xz\downarrow}h_{yz\downarrow}h_{yz\uparrow}+ \mathrm{H.c}),\nonumber
\end{eqnarray}
being $n_{\alpha}=n_{\alpha\uparrow}+n_{\alpha\downarrow}$, $n_{\alpha\sigma}=h^{\dagger}_{\alpha\sigma}h_{\alpha\sigma}$
where $\alpha$ indicates the orbital index $\{xz,yz\}$. $U$ ($U'$) represents the intra- (inter-)orbital Coulomb 
interaction and $J_H$ the Hund exchange coupling. 

The conduction bands are considered as non-interacting Hamiltonians,
\begin{equation} \label{cond-ham}
\tilde{H_{c}}=\sum_{k\alpha \sigma }\epsilon _{k\alpha}c_{k\alpha \sigma}^{\dagger }c_{k\alpha \sigma },  
\end{equation}
and the hybridization term that mixes both contribution is given by 
\begin{eqnarray} \label{mix-ham}
\tilde{H}_{mix}&=&\sum_{k\alpha \sigma }\left(V_{\alpha}c_{k\alpha \sigma}^{\dagger }d_{\alpha\sigma} + \mathrm{H.c}\right).  
\end{eqnarray}
For simplicity the analysis done below is restricted to $D=0$. The changes for the general case are straightforward.
While the Hamiltonian is explicitly invariant under spin rotations, the relation $U'=U-2J_H$ comes from 
spherical symmetry SO(3) of the Coulomb interaction including orbital degrees of freedom~\cite{aligia13,oles}.

The diagonalization of $\tilde{H}_{imp}$ within the two-hole subspace results in the triplet states 
\begin{eqnarray}\label{tripletes}
 &&h^{\dagger}_{xz\uparrow}h^{\dagger}_{yz\uparrow}\vert 0 \rangle,\nonumber\\
 &&\frac{1}{\sqrt{2}}(h^{\dagger}_{xz\uparrow}h^{\dagger}_{yz\downarrow}+h^{\dagger}_{xz\downarrow}h^{\dagger}_{yz\uparrow})
\vert 0\rangle,\\
&&h^{\dagger}_{xz\downarrow}h^{\dagger}_{yz\downarrow}\vert 0 \rangle\nonumber,
\end{eqnarray}
with energy $E_{T}=E_2=\epsilon_{xz}+\epsilon_{yz}+U'-J_H,$ together with the following two singlets, degenerate as a consequence of 
the $SO(3)$ symmetry of the interaction, 
\begin{eqnarray}\label{singletes_low}
&&\frac{1}{\sqrt{2}}(h^{\dagger}_{xz\uparrow}h^{\dagger}_{yz\downarrow}-h^{\dagger}_{xz\downarrow}h^{\dagger}_{yz\uparrow})
\vert 0 \rangle,\nonumber\\
&&\frac{1}{\sqrt{2}}(h^{\dagger}_{xz\uparrow}h^{\dagger}_{xz\downarrow}-h^{\dagger}_{yz\uparrow}h^{\dagger}_{yz\downarrow})
\vert 0 \rangle,
\end{eqnarray}
with energy  $E_{S_{low}}=\epsilon_{xz}+\epsilon_{yz}+U'+J_H$, and finally one excited singlet,
\begin{eqnarray}\label{singletes_ex}
\frac{1}{\sqrt{2}}(h^{\dagger}_{xz\uparrow}h^{\dagger}_{xz\downarrow}+h^{\dagger}_{yz\uparrow}h^{\dagger}_{yz\downarrow})
\vert 0 \rangle 
\end{eqnarray}
with energy $E_{S_{ex}}=\epsilon_{xz}+\epsilon_{yz}+U+J_H.$

The two-orbital model of Eq.~(\ref{2orb-ham}), neglecting the pair-hopping term [the last one in Eq.~(\ref{imp-ham})], 
has been studied by using NRG in the context of impurity~\cite{sakai89,izumida98,koyima03,deleo04,zhuravlev04, nishikawa12} 
and lattice models within the dynamical mean field theory~\cite{pruschke05}. Specifically, the work of Nishikawa and 
Hewson~\cite{nishikawa12} focuses on the role of Hund's interaction and as we shall see, some of our results agree with theirs.

In any case, since we are interested in retaining only the lowest triplet state in the configuration with two particles, we can 
take $U'=J_H$ with $J_H$ and $U$ large enough so that the singlets play no role and can be removed from the impurity Hilbert 
space. The condition $U'=U-2J_H$ is not satisfied but this only breaks the symmetry of irrelevant high-energy 
singlet states. 
We also neglect, as in previous works, the pair-hopping term (the double occupied states with both holes in the same 
orbital are excluded by large $U$). Then, the surviving two-particle states belong to the triplet with energy 
$E_2=E_T=\epsilon_{xz}+\epsilon_{yz}$, and the other relevant configuration has two one-particle doublets with energy 
$E_1=\epsilon_{\alpha}$. Note that the zero-particle state has a finite energy $E_0=0,$ but higher than the 
one-particle state energies; consequently, it can also be discarded in the study of the low energy physics of the $S=1$ impurity. 

The resulting accessible Hilbert space contains only the three components of the two-particle spin $S=1$ and the 
two one-particle doublets and, therefore, the Hamiltonian mixes these two configurations in identical form that the corresponding 
one in Eq.~(\ref{ham}). These assumptions highly simplify the use of NRG.

As mentioned in the introduction, we present calculations in the strong Hund's coupling limit, in which a $S=1$ ground state is 
screened by spin-$1/2$ electrons. Furthermore, the role of the anisotropy term, not included in previous studies,  
$DS^{2}_{z}$ is considered. We note that Ref.~\onlinecite{pruschke05} analyzes some aspects of this anisotropy contribution. 

\subsection{\textit{Non Crossing Approximation}}\label{sec_nca}

Although the non-crossing approximation \cite{bickers87} for fully screened models fails to accurately reproduce Fermi liquid 
relationships at zero temperature, it gives accurate results at finite and high excitation energies. For instance, the intensity 
and the width of the charge-transfer peaks of the spectral density 
(those which correspond to differences in energy between neighboring configurations, such as 
the dot level $\epsilon$ and $\epsilon+U$ in the simplest one-channel SU(2) impurity Anderson model) given by NCA 
were found~\cite{aligia15,fernandez18} to be in agreement  with other theoretical methods~\cite{pruschke89,logan98} 
and also with experiments 
in which a marked asymmetry in the intensity and width of the resonances for bias voltage $V\neq 0$ was observed, 
depending on the 
polarity of $V$~\cite{konemann06}. Furthermore, it has a natural extension to non-equilibrium conditions \cite{wingreen94} and 
it is especially suitable for describing satellite peaks away from the zero bias voltage~\cite{tosi15,dinapoli14,roura09}. 
In addition, the NCA Haldane shift~\cite{haldane78} (the renormalization of the bare $\epsilon_{\alpha}$ 
energy due to many body 
correlations) was found to be in agreement with the 
Haldane's prediction~\cite{tosi15,aligia15,fernandez18}. We remind the reader that the 
charge-transfer peaks as any other satellite peak in the spectral function are artificially 
broadened within NRG due to the 
logarithmic discretization of the conducting band~\cite{vaugier07} and hence the NCA solution,
which is free of this shortcoming became a useful alternative treatment. 
For the ($S=1/2$) one-channel case, the NCA reproduces well the scaling relations 
with temperature $T$ and bias voltage in the Kondo 
regime~\cite{roura10b}. 

Within the NCA framework, we rewrite the Hamiltonian in Eq.~(\ref{ham}) by using a pseudo-particle representation of the Hubbard
operators $\vert M_i >< M_j \vert \longrightarrow \hat{a}^{\dagger}_{M_i}\hat{a}_{M_j}$, which renders the model to the following 
form 
\begin{eqnarray} \label{nca-model}
H'&=&\sum_{M_{2}}(E_{2}+
D M_{2}^{2})\hat{a}^{\dagger}_{M_2}\hat{a}_{M_2} 
+\sum_{\alpha M_{1}}E_{\alpha}\hat{a}^{\dagger}_{\alpha M_{1}}\hat{a}_{\alpha M_{1}} + \nonumber\\
&+&\sum_{\nu k\alpha \sigma }\epsilon _{\nu k}c_{\nu k\alpha \sigma
}^{\dagger }c_{\nu k\alpha \sigma }+\\
&+&\sum_{M_{1}M_{2}}\sum_{\alpha\nu k\sigma }V_{\nu\alpha }\langle 1
M_{2}  |\frac{1}{2}\frac{1}{2}M_{1}\sigma \rangle 
(\hat{a}^{\dagger}_{M_2}\hat{a}_{\alpha M_{1}} c_{\nu k\alpha \sigma }\!+\!\mathrm{H.c.}).  \nonumber 
\end{eqnarray}
In addition,  the number of pseudo-particles should satisfy the constraint 
\begin{equation}
\sum_{M_{2}}\hat{a}^{\dagger}_{M_2}\hat{a}_{M_2}+
\sum_{\alpha M_{1}}\hat{a}^{\dagger}_{\alpha M_{1}}\hat{a}_{\alpha M_{1}}=1.
\end{equation}

The approximation makes use of an individual dynamics of each class of particles, which obeys the 
following self-consistent equations for the corresponding self-energies
\begin{eqnarray} \label{nca-selfenergies}
\Sigma_{\alpha}(\omega)&=&\frac{1}{\pi}\int~d\epsilon f(\epsilon) \Delta_{\bar{\alpha}}(\epsilon)
\left[G_{21}(\epsilon+\omega)+
 \frac{1}{2}G_{20}(\epsilon+\omega)\right],\nonumber \\
\Sigma_{2}(\omega)&=&\frac{1}{\pi}\int~d\epsilon f(-\epsilon) \sum_{\alpha}\Delta_{\alpha}(\epsilon)
 G_{\bar{\alpha}}(\omega-\epsilon),
\end{eqnarray}
where $\Delta_{\alpha}(\epsilon)=\pi V^{2}_{\alpha}\rho^{(c)}_{\alpha}(\epsilon)$ represents the hybridization 
of the impurity with the $\alpha$-channel of conduction electron of density 
$\rho^{(c)}_{\alpha}(\epsilon)$. The retarded Green function $G_{21}$ ($G_{20}$) takes into account the $\pm 1$ ($0$) 
components of the triplet, while $G_{\alpha}$ stands for the doublet of symmetry $\alpha$. The temperature is 
included within the Fermi function $f(\epsilon)$.

Details of the technique and its numerical evaluation can be found in the above mentioned references.

\section{Numerical results} \label{sec_numerical}
For a numerical resolution of the model at hands, constant and symmetric 
unperturbed conduction bands, $\rho^{(c)}_{\alpha}$, in 
the range $[-W,W]$ are considered.
Furthermore, without loss of generality (except for the 
magnitude of the current) we assume symmetric coupling to the leads, 
$V_{\alpha L}=V_{\alpha R}=V_{\alpha}/\sqrt{2}$, 
independent of energy, which implies a constant resonant-level width 
$\Delta_{\alpha}=\pi V_{\alpha}^2 \rho^{(c)}_{\alpha}$.  We also define $\Delta=\Delta_{xz}+\Delta_{yz}$

We restrict ourselves to the case in which both conducting band densities $\rho^{(c)}_{\alpha}$, hybridization
hoppings, $V_{\alpha}$, and the impurity levels $\epsilon_{\alpha}$ are degenerate by symmetry, 
inspired in the real situation of 
the Ni impurity in an Au chain (along the z-axis) for which the relations $\rho^{(c)}_{5d_{xz}}(\omega)=\rho^{(c)}_{5d_{yz}}(\omega)$, 
$V_{xz}=V_{yz}$, and $\epsilon_{xz}=\epsilon_{yz}$ hold.

We define the unit of energy ($W=1$) such that $\Delta=0.01$. We also take $U'=J_{H}=1000$ in the 
rest of the paper, unless otherwise stated. 
This choice displaces to very high energy all excited states of the $d^8$ configuration, leaving only 
the triplet states $|1M_2 \rangle$ at low energies. As explained in Section \ref{sec_nrg}, this choice 
allows us to represent the model Eq. (\ref{ham}) in a form suitable for the NRG code.
Except for the subsection~\ref{sec_anisotropy}, in the rest of the paper we take  
an anisotropy $D=0$.

For the cases in which the system is in the Kondo regime, that is for $\Delta$ much smaller than the other
bare energy scales, we define the Kondo temperature $T_K$ from the equilibrium conductance in the following way,
$G(T_K)=G(T\rightarrow0)/2$. This definition of $T_K$ gives values of the same order of magnitude than the 
corresponding one obtained from the half width at half maximum of the Kondo resonance 
in the spectral density~\cite{diego}.
Regarding the bare orbital energies, we employ the notation $\epsilon_d=E_2-E_1$. 

\subsection{\textit{Entropy and occupancy}}\label{sec_entropy_occupancy}
\begin{figure}[ht]
\includegraphics*[width=0.85\columnwidth]{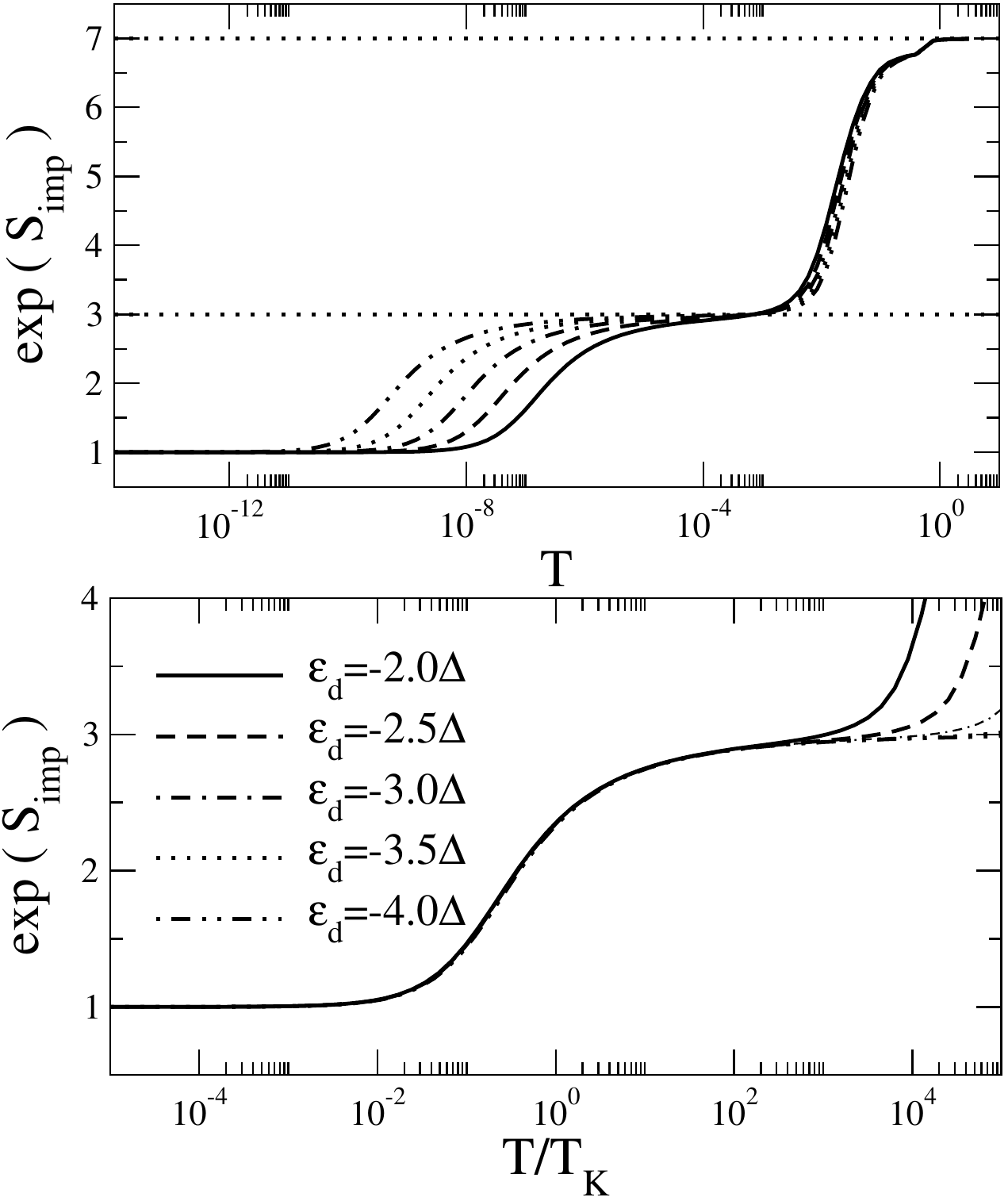}
\caption{Top panel: NRG impurity contribution to the entropy as a function of temperature, $S_{imp}(T)$,
for the model of Eq.~(\ref{2orb-ham}). Lower panel: Same data in units of $T/T_K,$ being $T_K=6.7 \times 10^{-7},
~ 1.7 \times 10^{-7},~ 4.0 \times 10^{-8}, ~ 9.2\times 10^{-9}, ~2.1 \times 10^{-9}$ for 
$-\epsilon_d/\Delta=2$, $2.5$, $3$, $3.5$, $4$, respectively.} 
\label{fig:entropia}
\end{figure}

We start our discussion of the numerical results of the model of Eq.~(\ref{ham}) by analyzing the NRG results for 
the impurity contribution $S_{imp}(T)$ to the total entropy as a function of temperature.
The top panel of Fig.~\ref{fig:entropia} shows the calculated $S_{imp}(T)$ for four different set of parameters, 
$\epsilon_d=\{-4, -3.5, -3, -2.5, -2\}\Delta$ as a function of temperature. The case $\epsilon_d=-2\Delta$ correspond to 
the {\it ab-initio} parameters for the Ni impurity in an O-doped Au chain~\cite{dinapoli15} (black solid line).

At large enough temperatures, $e^{S_{imp}}$ saturates at the value imposed by the 
dimension of the local Hilbert space  $g=7$ 
given by the three components of the triplet and the 4-fold degenerate states corresponding to the two doublets. 
As the temperature is lowered, an intermediate plateau can be observed in which $e^{S_{imp}} \simeq  3,$ due to the 
triplet. This is the local-moment regime characterized by the fact that the charge fluctuations are frozen. It corresponds to 
the Kondo limit of the model, obtained from a Schrieffer-Wolff transformation in Ref.~\onlinecite{dinapoli15}, in which only 
spin fluctuations are present. 

As expected for the symmetry of the model, in which two spin-$1/2$ conduction bands coherently screen the total impurity spin 
$S=1$, when the temperature falls under $T_K$, the system enters the strong-coupling regime, and the value of $e^{S_{imp}}$ 
tends to one, corresponding to the Fermi-liquid non-degenerate Kondo ground state.

\begin{figure}[ht]
\includegraphics*[width=0.75\columnwidth]{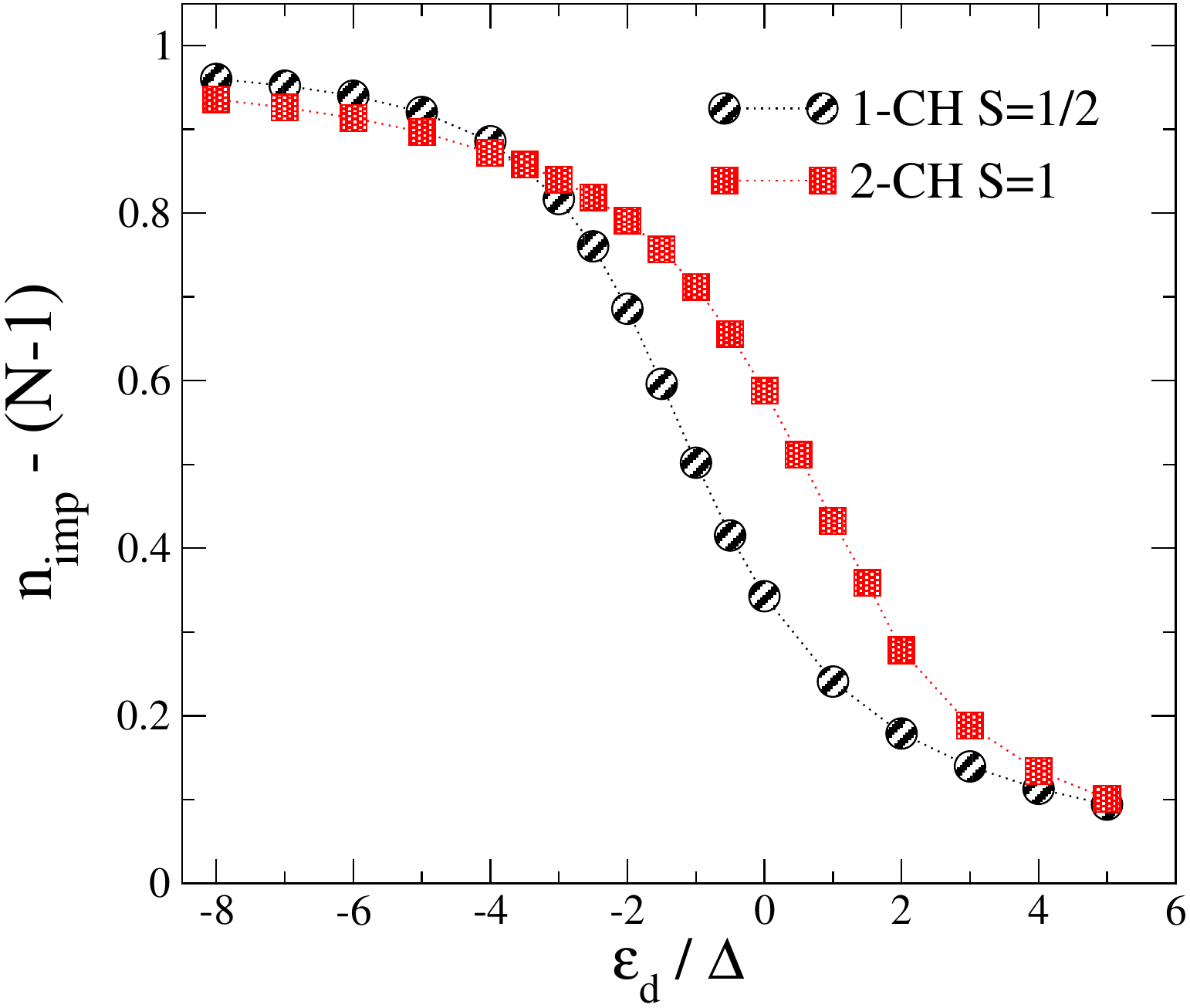}
\caption{(Color online) NRG total impurity occupancy, $n_{imp}-(M-1),$ as a function of $\epsilon_d$, being $M$ the number of channels. 
Squares indicate the results for the model of Eq.~(\ref{2orb-ham}), $M=2$. The circles correspond to the impurity population 
for the spin $s=1/2$ infinite $U$-limit one-channel Anderson model where $M=1$.}
\label{fig2:ocupacion}
\end{figure}

In the lower panel of Fig.~\ref{fig:entropia} we show the same data but with temperatures scaled by the corresponding Kondo ones. 
It is clear that, within the local-moment and strong-coupling regimes, the system displays universality and all curves have the 
same temperature dependence when the latter is expressed in units of $T_K$, being this the only relevant scale of the model.

Note that, for the {\it ab-initio} parameters, there is also an extended range of temperatures, of about 7 decades, 
$10^{-5}\lesssim T/T_K\lesssim 10^{2}$, in which $e^{S_{imp}}$ falls on top of the others corresponding for more negative values of 
$\epsilon_d$. This is a common feature of the regime of parameters $\epsilon_d \ll -\Delta$ for which the Kondo model is valid. 
However, $\epsilon_d=-2\Delta$ would seem not negative enough to suppress charge fluctuations. Indeed, the relation 
$\epsilon_d=-2\Delta$ for the case of the well studied one-channel spin-$1/2$ Anderson impurity characterizes the mixed valence 
regime of the model~\cite{roura10b}. To clarify this point, we compared the NRG results for the total impurity occupancy $n_{imp}$ 
as a function of the energy $\epsilon_d$ in the present case of the 2-channel $S=1$ model and the one corresponding to the 
one-channel spin-$1/2$ impurity model. The result is shown in Fig.~\ref{fig2:ocupacion}.

Remarkably, the mixed valence regime in the case of the 2-channel $S=1$ is strongly suppressed in the range of negative
values of $\epsilon_d$. In fact, when $-\epsilon_d/\Delta=1$ the impurity is near 70\% occupied as compared with near 50\%
in the case of the ordinary one-channel spin-$1/2$ model. From this result, we conclude  that the realistic parameters representing 
the Ni impurity in the Au chain correspond to a description of the system within its Kondo regime.  
This point will be discussed further in the following subsections. 

\subsection{\textit{Spectral density}}\label{sec_spectral_density}
\begin{figure}[ht]
\includegraphics*[width=0.75\columnwidth]{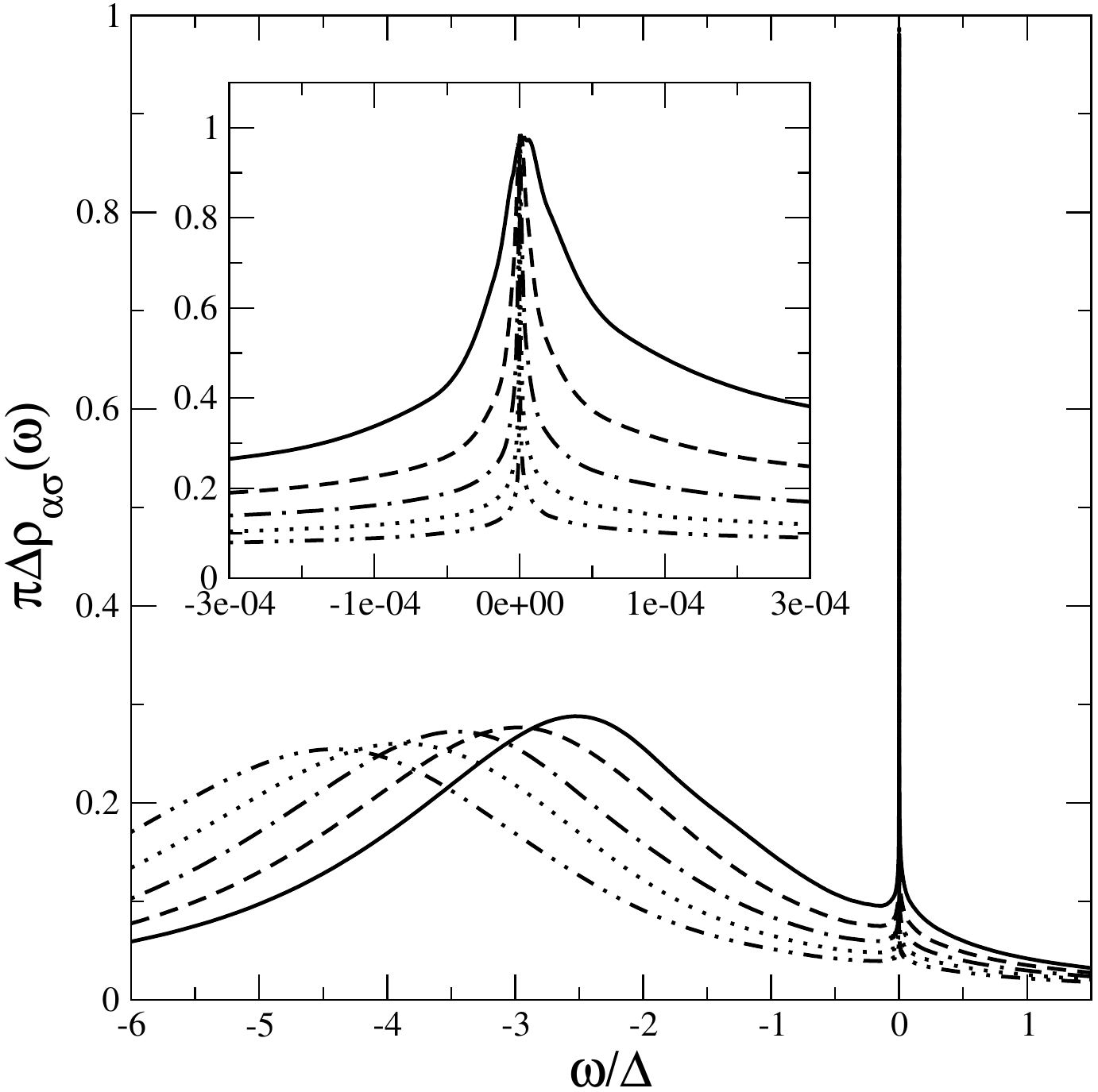}
\caption{NRG impurity spectral density as a function of frequency for the same set of values of $\epsilon_d$ and other parameters as 
in Fig.~\ref{fig:entropia}.}
\label{fig3:spectral-nrg}
\end{figure}

In Fig.~\ref{fig3:spectral-nrg} we show the NRG impurity spectral density per channel and per spin $\rho_{\alpha\sigma}$ 
as a function of the frequency for several values of the bare energy level $\epsilon_d$ and at sufficiently low temperature 
as compared with the Kondo one for each $\epsilon_d$. The resulting spectral function is quite similar to the corresponding 
one in the case of the one-channel infinite $U$-limit spin-$1/2$ Anderson impurity. Indeed, there is only one charge transfer 
peak located near the bare energy $\epsilon_d=E_2-E_1$ which indicates the energy needed to put a second hole (-electron) in 
the impurity to form one of the triplet states. Furthermore, the narrow Kondo peak at the Fermi level has a width of the order 
of $T_K$ (visible in the inset of the figure) and its intensity is imposed by the usual Friedel sum rule, 
$\rho_{\alpha\sigma}(0)=\frac{1}{\pi\Delta}{\rm sin}^{2}(\frac{\pi n_{\alpha}}{2})$, where $n_{\alpha}$ is the total 
population of the $\alpha$-level.

However, there are non-trivial differences between the spectral density of the present model and one-channel infinite $U$-limit 
spin-$1/2$. The first one is related to the reduced width of the Kondo resonance, whose discussion will be given in detail in the 
next subsection. Here we focus on the manifestation of the many-body interactions in the charge-transfer peak. Specifically,
we examine its width and position. To this purpose, we employ the NCA results for the spectral density. Although the NCA does not 
provide accurate results for the low energy physics of the model at hands, in particular $T_K$ is found to be overestimated as we 
will show in the appendix~\ref{app_kondo_nca}, it is especially suitable for the study of the charge transfer peak. We remind the 
reader that the logarithmic discretization of the conduction band produces an artificial broadening of the charge transfer peaks within 
the NRG procedure~\cite{vaugier07}. Therefore, we use NCA when analyzing high energy scales. \\

\begin{figure}[ht]
\includegraphics*[width=0.75\columnwidth]{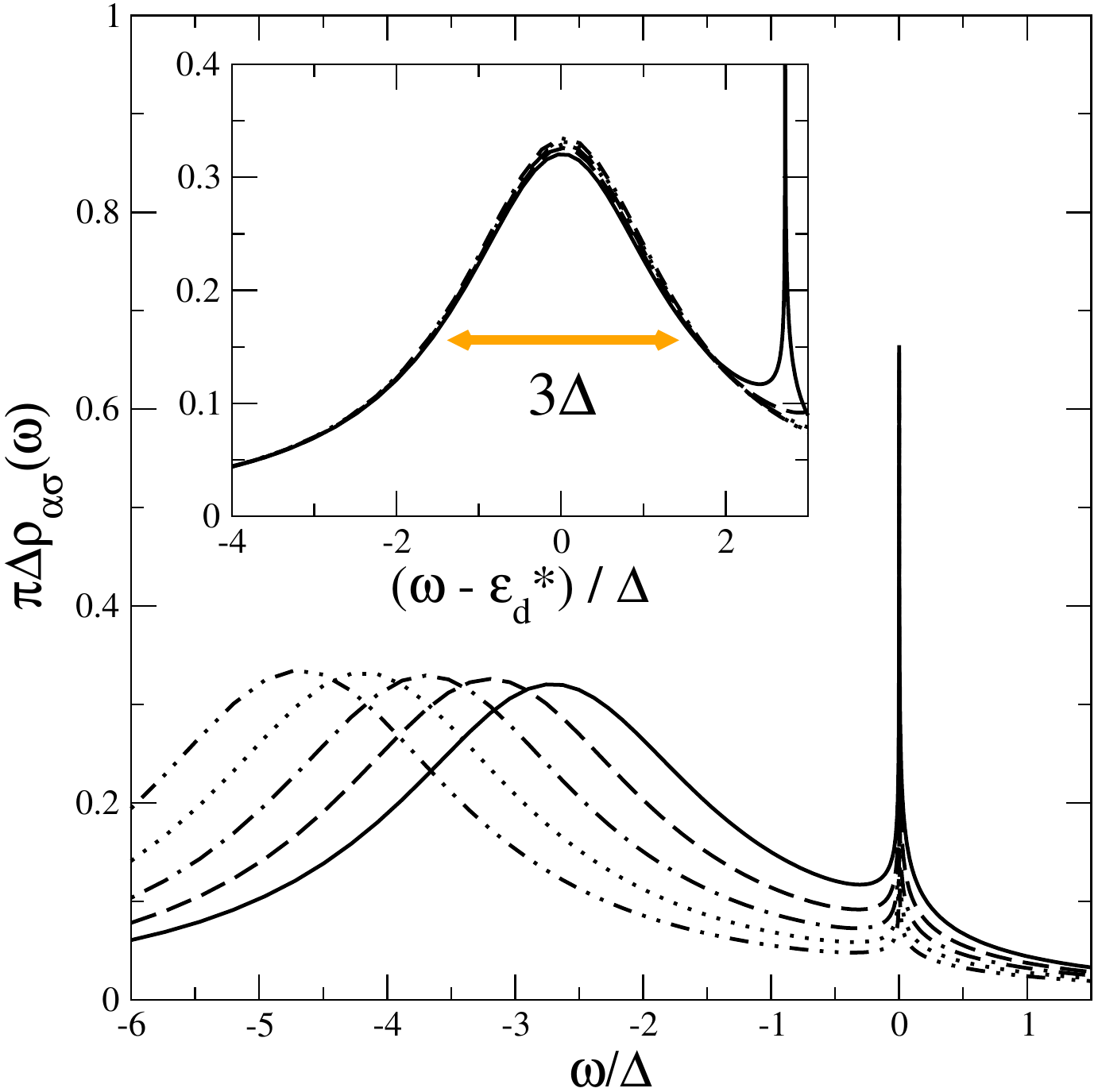}
\caption{Same as Fig.~\ref{fig3:spectral-nrg} calculated with NCA. The inset shows
the charge-transfer peaks shifted by $\epsilon^{\ast}_{d}$ (see Eq.~\ref{haldane-shift}).}
\label{fig4:spectral-nca}
\end{figure}

\textit{Position and width of the charge transfer peak}.
In Fig.~\ref{fig4:spectral-nca} we present the NCA results for the spectral density for the same set of parameters as in 
Fig.~\ref{fig3:spectral-nrg}.

\vspace{0.5cm}

\textit{Renormalization of the bare energy $\epsilon_d$: The Haldane shift}.
By means of scaling theory, Haldane~\cite{haldane78} has showed that for the asymmetric spin $s=1/2$ one-channel Anderson model 
($U\gg\vert\epsilon_d\vert,\Delta$), the renormalized value of the level $\epsilon_d$ is given by 
$\epsilon^{\ast}_{d}=\epsilon_d + (\Delta/\pi){\rm ln}(W/\Delta)$. As we have mentioned in section~(\ref{sec_nca}), 
the NCA has proved to be capable of providing accurate results for such energy shift~\cite{tosi15,aligia15,fernandez18}.
The sign and the main features of the dependence with $\Delta$ of the Haldane shift 
can be understood by a simple argument. 
Due to the hybridization with the conduction band, the bare energies of the empty and the single occupied states are renormalized. 
While the empty state is mixed with both single occupied states and, consequently, its energy is lowered by an amount 
proportional to $2 \Delta$, a single occupied state can only be mixed with the empty state, so its bare energy $\epsilon_d$ 
is reduced an amount $\propto \Delta.$ As $\epsilon_d^\ast$ is the energy difference between both renormalized energies, 
we have $\Delta\epsilon \equiv \epsilon_d^\ast - \epsilon_d \propto \Delta$. Of course, the logarithmic term 
in $\Delta \epsilon = (\Delta/\pi){\rm ln}(W/\Delta)$ can only be obtained through the scaling process. 
This approach is easy to generalize to the SU($N$) impurity Anderson model for valence fluctuations between
the configurations with zero and one localized particles~\cite{fernandez18}.

In the case of the model describing a Ni impurity and following Haldane's approach, starting from the triplet state 
$\vert 11 \rangle = {h}_{xz\uparrow}^{\dagger}{h}_{yz\uparrow}^{\dagger}\vert 0 \rangle$ there are two different processes 
with hybridization $V$ that connect this state with one containing only one hole, implying that its renormalized energy goes 
down an amount $\propto 2\Delta$. On the other hand, starting from a given 
state $\vert \alpha \uparrow \rangle $ and assuming $J_H \rightarrow \infty$ there are also two processes, one to the state 
$\vert 11 \rangle$ with hybridization $V$ and other one to the state $\vert 10 \rangle=\frac{1}{\sqrt{2}}(h^{\dagger}_{xz\uparrow}h^{\dagger}_{yz\downarrow}+h^{\dagger}_{xz\downarrow}h^{\dagger}_{yz\uparrow})
\vert 0 \rangle$ with hybridization $V/\sqrt{2},$ in such a way that the bare energy of the one-hole state goes down $\propto 3\Delta/2$. 
This energy gain is lower than that corresponding to the triplet states, and so we expect $\Delta\epsilon \propto -\Delta/2$.
In fact, using scaling arguments similar to those of Ref.~\onlinecite{haldane78}, the renormalized energy $\epsilon^{\ast}_{d}$ 
becomes
\begin{equation}\label{haldane-shift}
 \epsilon^{\ast}_{d}=\epsilon_d - (\Delta/2\pi){\rm ln}(W/\Delta).
\end{equation}
Note that in comparison to the known one-channel $s=1/2$ case, the shift 
$\Delta\epsilon$ has the opposite sign and its magnitude is reduced by a factor 1/2. 

\begin{figure}[ht]
\includegraphics*[width=0.85\columnwidth]{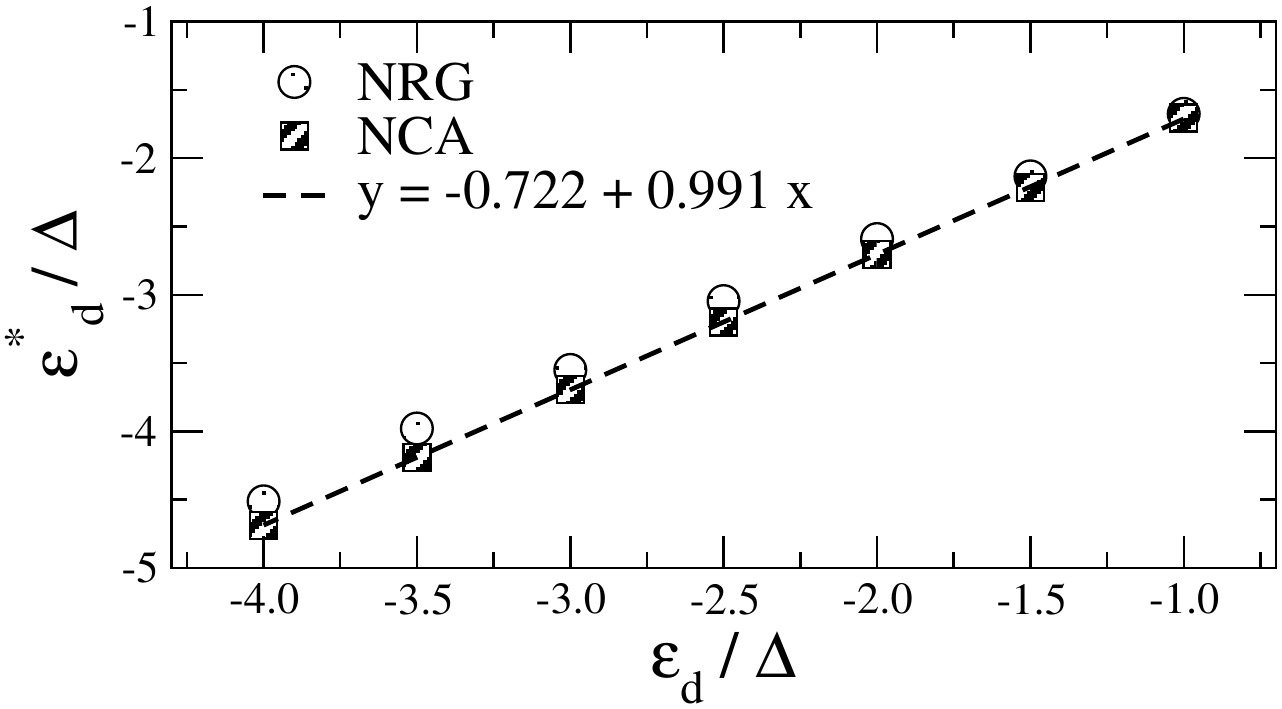}
\caption{Renormalized position of the charge-transfer peak, $\epsilon^{\ast}_{d}$, 
as a function of $\epsilon_d$. Other parameters as in Fig.~\ref{fig:entropia}.}
\label{fig5:shift_ancho}
\end{figure}

We calculate $\epsilon^{\ast}_{d}$ as the energy position of the maximum in the charge-transfer peak of the spectral density 
in both, NRG and NCA.  Fig.~\ref{fig4:spectral-nca} explicitly shows the shift towards negative energies.
In Fig.~\ref{fig5:shift_ancho} the values of $\epsilon^{\ast}_{d}$ are shown as a function of $\epsilon_d$ in units of $\Delta$.
Both techniques display a linear behavior, however with NCA we obtain a remarkably better well defined linear function 
(correlation coefficient 0.999) with a slope close to $1$ and a constant of $-0.722$ in agreement with the second term of the 
r.h.s. of Eq.~\ref{haldane-shift}, $-0.733$.

\vspace{0.5cm}

\textit{Width of the charge-transfer peak}. 
The inset of Fig.~\ref{fig4:spectral-nca} shows a detail of the charge-transfer peak shifted by $\epsilon^{\ast}_{d}$. 
We remind the reader that the half-width of this peak in the case of the one-channel $s=1/2$ case is found to be 
$2\Delta$, where $\Delta$ corresponds to the one-body broadening already present 
in the non-interacting model~\cite{pruschke89}. The prefactor which in general is $N$ for the SU($N$) case ~\cite{fernandez18} has its origin in 
effects of the interaction.

In the present case the half-width of the charge-transfer peak is $3\Delta/2$. 
We trace back this difference with the following qualitative argument: 
the half-width $2\Delta$ in the one-channel $s=1/2$ model reflects the two processes by which the excited empty state 
is connected to the single-occupied ground state with hybridization $V$~\cite{fernandez18}.
On the other hand, for the $s=1$ model, the excited one-hole states are connected to the $S_z=1$ (or $S_z=-1$) component 
of the ground state with hybridization $V$, and to the $S_z=0$ component with hybridization $V/\sqrt{2}$. As a consequence, the half-width is now 
$3\Delta/2$.

\subsection{\textit{Kondo temperature and Kondo resonance narrowing effect}}\label{sec_reduction_tk}
The Kondo temperature is undoubtedly the most relevant energy scale in the Kondo phenomena simply because it represents a 
universal scale, in terms of which all physical properties follow a given dependence as a function of $T/T_K$ without being 
affected by the other parameters of the model. Therefore, it is always desirable to have an analytical expression for such a scale. 

Previous studies on the basis of NRG calculations have found that the introduction of Hund's
coupling into the Anderson model causes an exponential reduction in the Kondo temperature~\cite{nishikawa12,pruschke05}. 
Particularly, our model assumes an infinite Hund's coupling and belongs to the same class as the one studied by Nevidomskyy and 
Coleman in Ref.~\onlinecite{nevidomskyy09} by means of scaling arguments. Applied to our case, their main result was the existence of a 
factor 1/2 in the exponent of the expression of $T_K$ for the full screened spin $S=1$ model in comparison with the corresponding one 
for the spin $s=1/2$. In general, all approaches agree in the exponential dependence of the Kondo scale, given by 
${\rm exp}{\left(\frac{\pi\epsilon_d}{\Delta}\right)}$ instead of ${\rm exp}{\left(\frac{\pi\epsilon_d}{2\Delta}\right)}$ 
of the usual one-channel spin $s=1/2$ case.  
Therefore, a relation of the form $T^{S=1}_{K}\sim \left( T^{s=1/2}_{K} \right)^{2}$ is expected being $ T^{s=1/2}_{K}=
\sqrt{W\Delta}~{\rm exp}{\left(\frac{\pi\epsilon_d}{2\Delta}\right)}$. Note that the latter can be obtained from the 
Haldane~\cite{haldane78}  expression $T^{s=1/2}_{K}=\Delta~{\rm exp}{\left(\frac{\pi\epsilon^{\ast}_{d}}{2\Delta}\right)}$ using 
$\epsilon^{\ast}_{d}=\epsilon_d + (\Delta/\pi){\rm ln}(W/\Delta)$.
Following similar arguments, using $T^{S=1}_{K}=\Delta~{\rm exp}{\left(\frac{\pi\epsilon^{\ast}_{d}}{\Delta}\right)}$
and the renormalized
level position given by Eq.~(\ref{haldane-shift}) we obtain
\begin{equation}\label{kondo-scale-spin-1}
T^{S=1}_{K} = c\sqrt{\Delta/W^3} \left( T^{s=1/2}_{K} \right)^{2} = c\sqrt{\Delta^3/W}~e^{\frac{\pi\epsilon_d}{\Delta}},
\end{equation}
being $c$ a constant of the order of one.

\begin{figure}[ht]
\includegraphics*[width=0.85\columnwidth]{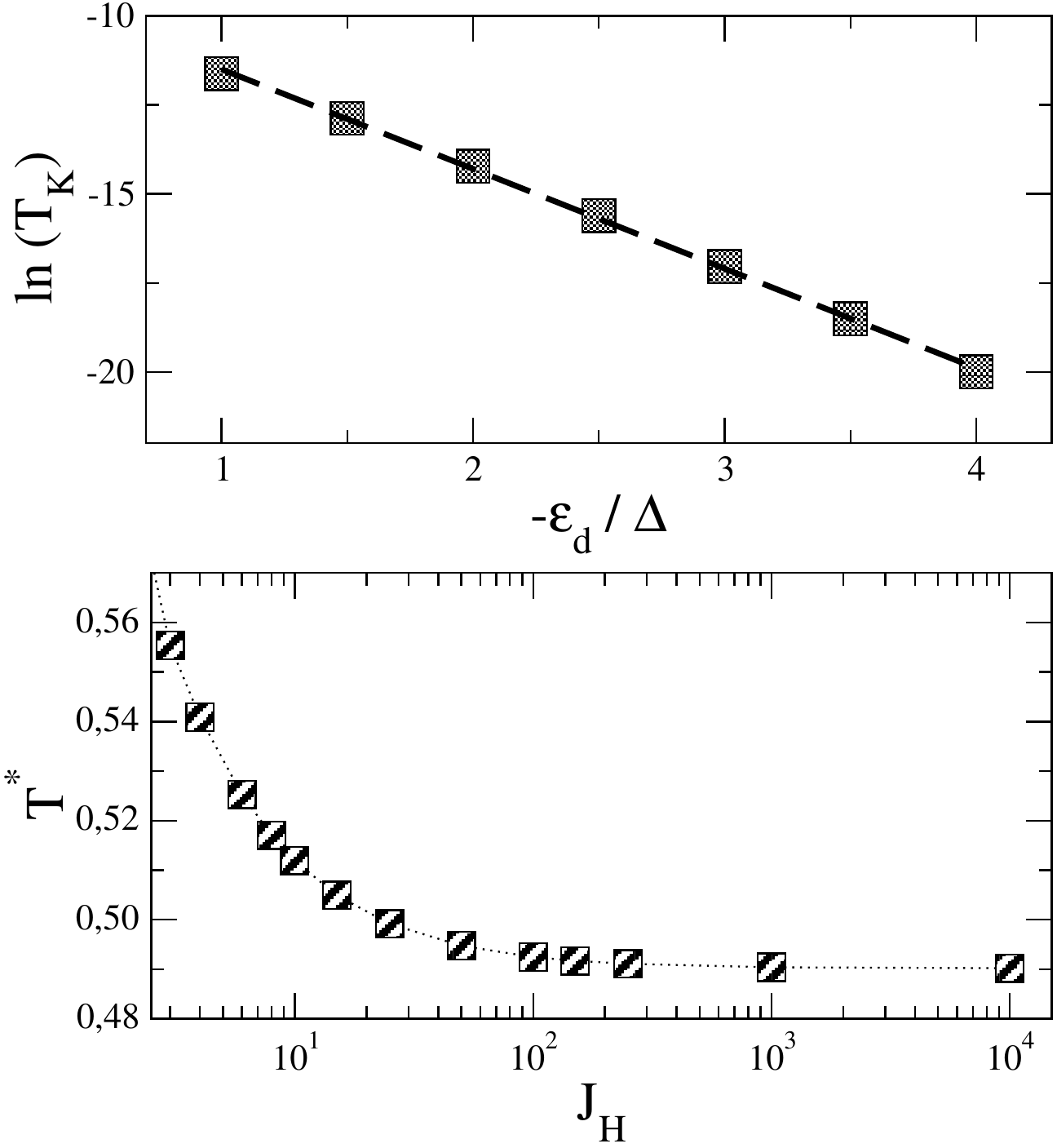}
\caption{Top panel: The squares indicate the NRG Kondo temperature as a function of 
$\vert \epsilon_d\vert/\Delta$ calculated for $\vert J_H \vert = 1000 W$. Dashed line:
linear fitting giving ${\rm ln}(T_K)=-8.70-2.80\vert \epsilon_d\vert/\Delta$ with a correlation coefficient of $0.999$. 
Lower panel: $T^\ast=T_{K}(J_H)/\sqrt{\Delta/W^3} ( T^{s=1/2}_{K} )^{2}$ as a function of $J_H$ for
$\vert \epsilon_d\vert/\Delta=3$.}
\label{fig6:tk}
\end{figure}

Our numerical NRG data for the Kondo temperature in case of large enough $J_H$, in such a way that the local moments become locked 
into a spin $S=1,$ confirm this exponential dependence. In the top panel of Fig.~(\ref{fig6:tk}) we show $T_K$
(obtained from the conductance as described at the beginning of this Section)
as a function of $\vert \epsilon_d\vert/\Delta,$ together a linear fit of the data which exhibits a slope that differs
from the factor $\pi$ in less than 10\%.

Regarding the dependence of $T_K$ with $J_H$ , in the lower panel of Fig.~(\ref{fig6:tk}) we show the calculated Kondo 
temperature in units of  $\sqrt{\Delta/W^3} \left( T^{s=1/2}_{K} \right)^{2}$ for a fixed value of $\vert \epsilon_d\vert/\Delta=3$.
It can be observed that the relation in Eq.~(\ref{kondo-scale-spin-1}) is satisfied in the asymptotic behavior $J_H\rightarrow \infty,$ 
being the constant $c=0.49\sim1/2$. In fact, we have verified that this constant varies between $0.5$ to $1.0$ for the whole set of 
values of $\vert \epsilon_d\vert/\Delta$ presented in top panel of Fig.~(\ref{fig6:tk}).

\subsection{Electrical conductance}\label{sec_conductance}
Since the ground state of model Eq.~\eqref{ham} for $D=0$ is a Fermi liquid, transport measurements through the proposed nanowire should exhibit 
universal behavior at low enough temperatures, for parameters that drive the system inside the Kondo regime. 
However, as we have shown in previous sections, the universal dependence of the observables, for instance as a function of temperature, 
is expected to be different from the well known spin $s=1/2$ case.  Here, we analyze the NRG results for the equilibrium electrical 
conductance, $G(T)$. The temperature dependence of the conductance through the Ni atom depends on the total impurity spectral function
$\rho(\omega) = \sum_{\alpha\sigma} \rho_{\alpha\sigma}(\omega),$
and it can be calculated from the following expression~\cite{wingreen94}
\begin{equation}\label{conductance}
 G(T) =  G_0 \frac{\pi\Delta}{2}\int ~d\omega (-f'(\omega)) \rho(\omega),
\end{equation}
where $G_0=2e^2/h$ is the quantum of conductance. Note that from the expected Friedel sum rule at zero temperature, 
$\sum_{\alpha\sigma}\rho_{\alpha\sigma}(0)\sim\frac{4}{\pi\Delta}$, the maximum value of the conductance should be $G(T\rightarrow0)=2G_0$, 
twice of the usual one channel case. We remind the reader that this is not always the case in two-channel models. For instance, in the 
overscreened $s=1/2$ two-channel case, the maximum value of the total conductance is found to be $G_0/2$~\cite{dinapoli13,dinapoli14}. 
The present result is a consequence of the Fermi liquid nature of the ground state, due to the full screening of the impurity spin.

\begin{figure}[ht]
\includegraphics*[width=0.85\columnwidth]{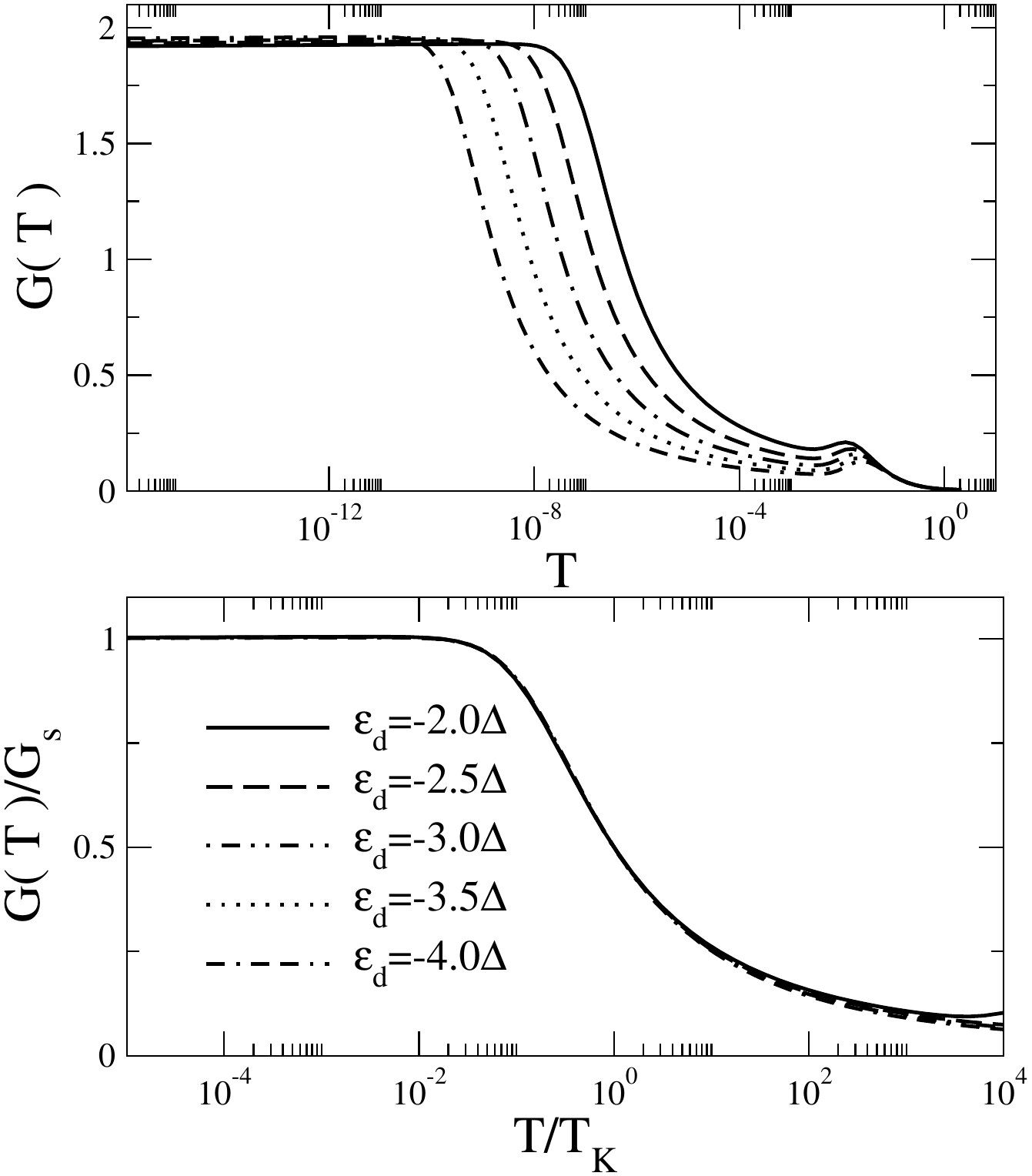}
\caption{Top panel: Equilibrium conductance $G(T)$ in units of $2e^2/h$ as a function of temperature. 
Lower panel: The same data that in the top panel in units of $T/T_K$ (see caption of Fig.~\ref{fig:entropia} for the values of $T_K$) 
and normalized by the saturated conductance $G_s=G(T\rightarrow 0)$}.
\label{fig7:conductance}
\end{figure}

In the top panel of Fig.~\ref{fig7:conductance}, we show the NRG results for $G(T)$ that correspond to the model of Eq.~(\ref{2orb-ham}) 
for several values of $\epsilon_d$. The lower panel displays the same data as in the top panel with the temperature scaled by
the corresponding Kondo one.  As expected, for temperatures $T\lesssim T_K$ the whole set of curves follows the same dependence in units of 
$T/T_K$, which confirm that $T_K$ is the only one relevant energy scale of the model within the Kondo regime. Furthermore, the 
case $-\epsilon_d/\Delta=2$ also follows the universal dependence reinforcing the conclusion that, for the parameters representing the Ni 
impurity in the O-doped Au chain, the system lies within the Kondo regime.

Regarding the dependence of $G$ with $T/T_K$, it is well known that the empirical expression 
\begin{equation}
G(T)=\frac{G_{s}}{\left[ 1+\left( 2^{1/s}-1\right) \left( T/T_{K}\right)
^{2}\right] ^{s}},  \label{ley_empirica}
\end{equation}
with $s=0.22$ and $G_s$ the conductance at temperature $T=0$, matches not only experimental results but also NRG calculations in the case of 
the spin $s=1/2$ one-channel Anderson model~\cite{goldhaber98,costi00}. 
A similar scaling law has been used to fit experimental data and one-channel NRG results 
for the resistivity due to magnetic impurities with 
spin $S=1/2$, 1 and 3/2 in Fe and Ag by Costi {\it et al}~\cite{costi09}. 
Although the scaling function in this case is different from 
that of $G(T)$ both coincide for $T \ll T_K$~\cite{cz}. Similar scaling functions were used to fit $G(T)$ for the 
underscreened Kondo model~\cite{parks10,roch09}. 
Here, we show that also for our model the scaling is noticeable different that in the $s=1/2$ case, 
in spite of being a fully compensated Kondo effect.

\begin{figure}[ht]
\includegraphics*[width=0.85\columnwidth]{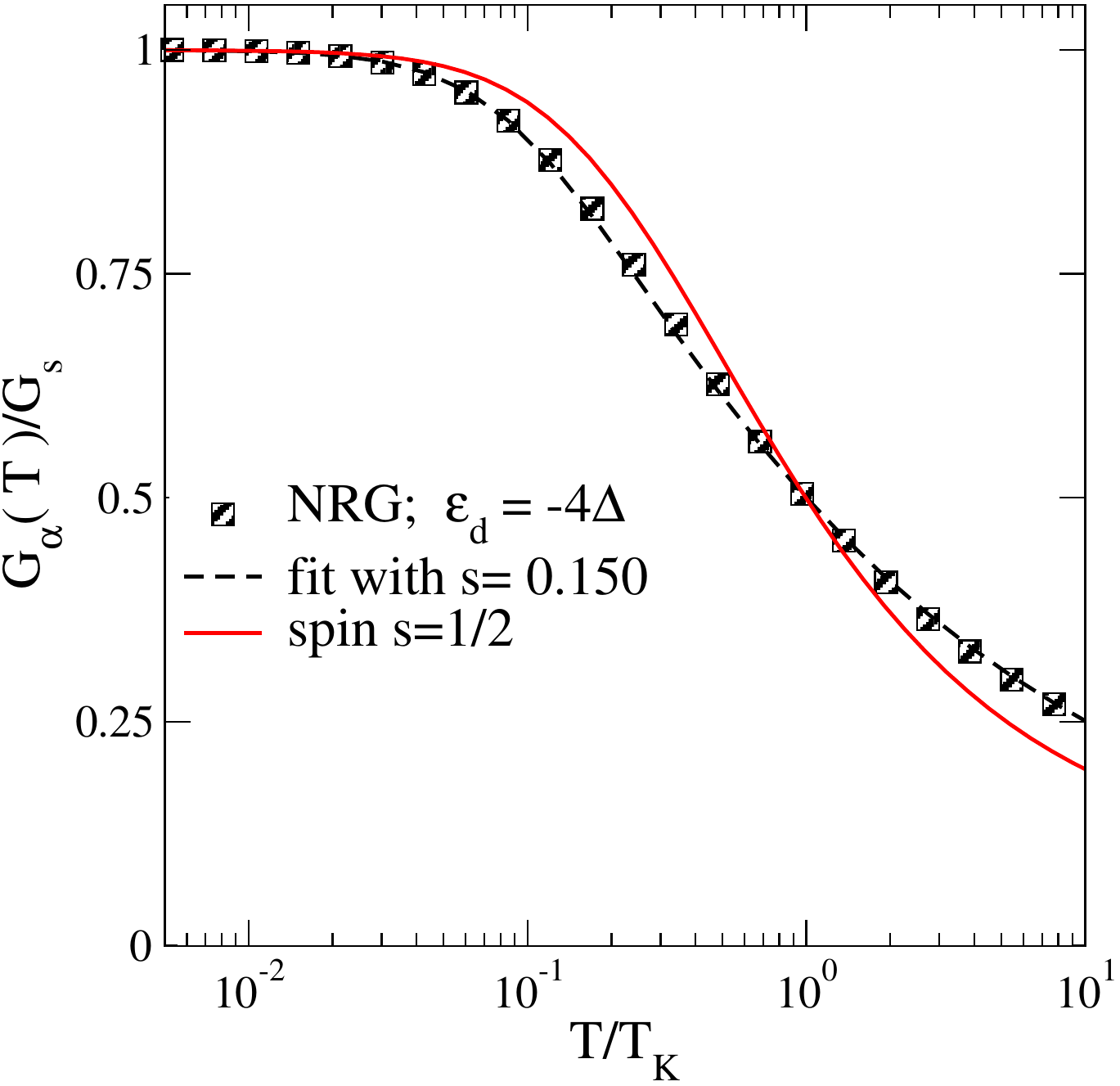}
\caption{(Color online) Squares: NRG data for $G_{\alpha}(T)=\sum_{\sigma}G_{\alpha\sigma}(T)$ in units of its maximum 
$G_s$ as a function of $T/T_K,$ for $\epsilon_d=-4\Delta$. Dashed black line: fitting of numerical data with expression
Eq.~(\ref{ley_empirica}) with $s=0.15$. Solid red line: Eq.~(\ref{ley_empirica}) with $s=0.22$.}
\label{fig8:universal_g}
\end{figure}

Fig.~(\ref{fig8:universal_g}) displays the NRG results for the total conductance per channel, $G_{\alpha}(T)=\sum_{\sigma}G_{\alpha\sigma}(T)$, 
in units of its maximum $G_s$ as a function of $T/T_K$ for a selected value of $\epsilon_d$ well inside the Kondo regime. 
The red solid line indicates the temperature dependence given by Eq.~(\ref{ley_empirica}) for the spin $s=1/2$ 
model. As it can be seen, the numerical data corresponding to the full screening of the spin $S=1$ appreciably deviates
from the latter. 

Our results indicate that the relation given in Eq.~(\ref{ley_empirica}) is still valid in the present case but with a
different factor $s=0.15$.  We have verified that similar coefficients, $0.15 < s < 0.17$, appear when the formula is applied to other,  
negative enough, values of $\epsilon_d$. 
Curiously, a similar exponent $s=0.16$ was found for the fit of the resistivity due to $S=1$ Kondo impurities in the 
underscreened case. 
Empirical formulas like (\ref{ley_empirica}) have been shown to be a very useful tool in order 
to discern the spin of impurities in experimental underscreened Kondo systems~\cite{parks10,roch09}, 
and so, the good agreement between the NRG $G(T)$ 
and Eq.~(\ref{ley_empirica}), would allow to identify $S=1$ fully screened systems.

\subsection{Role of the single-ion magnetic anisotropy}\label{sec_anisotropy}
A key ingredient in magnetic nanosystems is the presence of rather large single-ion magnetic 
anisotropies $DS_z^2$, due to the enhanced spin-orbit coupling brought about by the lower symmetries 
than in bulk systems~\cite{otte08,oberg13,heinrich15,hiraoka17,jacob18}.
In fact, cluster and {\it ab-initio} calculations~\cite{dinapoli15} yields an appreciable positive 
$D \sim 8.5$ meV for the Ni atom embedded into the O-doped Au chains. 

Very recently, working with the same model of Eq.~(\ref{2orb-ham}), we have uncovered a topological quantum 
phase transition between two Fermi liquids as a function of the magnetic anisotropy~\cite{blesio18}. 
For $D < D_c \simeq (2-3) T_K^0$ ($T^0_K$ is the Kondo temperature for $D=0$), the impurity is fully Kondo 
screened as in the $D=0$ case, with a Kondo temperature that, surprisingly, scales as a power law of $D$, 
\begin{equation}
 T_K(D) \propto T_K^0 \left(\frac{D_c-D}{D_c}\right)^2.
\end{equation}
close to $D_c$~\cite{blesio18}. On the hand, for $D > D_c$ the impurity spin is quenched by the anisotropy, 
giving rise to a topological non-trivial Fermi liquid ground state, 
with a non-zero Luttinger integral $I_L$~\cite{blesio18,curtin18}. 
Due to this fact the system cannot be adiabatically connected to a 
non-interacting one. Therefore we have named it a non-Landau Fermi liquid~\cite{blesio18}. At the critical 
point, the system exhibits several non-Fermi liquid signatures of the two-channel $s=1/2$ Kondo (2CK) effect. 

We remind the reader that in a Fermi liquid, as in a simple pure metal, the life time of the quasiparticles
for small excitation energy $\omega$ from the Fermi energy, scales as $\omega^{-2}$ at zero temperatures. 
Landau postulated that due to restrictions of phase space imposed by Pauli principle, the same behavior should
take place in interacting systems. However for some strongly interacting systems, like the overscreened Kondo models,
this picture breaks down and the quasiparticles have finite lifetime even at $\omega=0$. An intermediate
case are the marginal Fermi liquids (corresponding to underscreened Kondo models) in which
the lifetime is infinite at $\omega=0$ but has a non analytical dependence on $\omega$~\cite{mehta05,logan09}
In our case for $D \neq D_c$, the system is a Fermi liquid. Until recently, the natural expectation was that a Fermi
liquid, like a non-interacting system was characterized by $I_L=0$, but as shown first by 
Curtin {\it et al.}~\cite{curtin18}, some interacting system (such as ours for $D > D_c$) 
behave as Fermi liquids with $I_L \neq 0$.
$I_L$ has a topological character and can only have discrete values~\cite{blesio18,seki17}.
\begin{figure}[ht]
\includegraphics*[width=1.0\columnwidth]{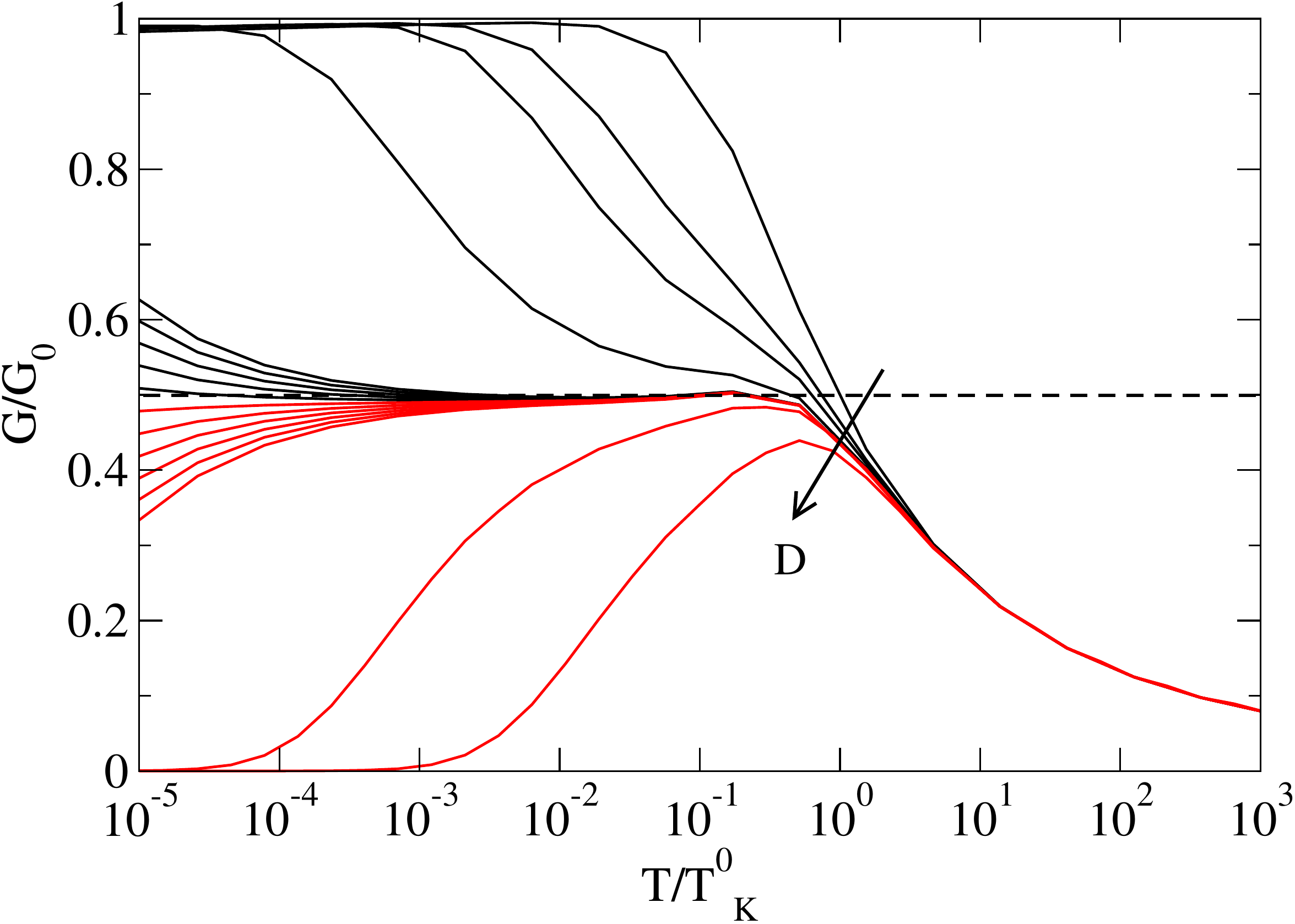}
\caption{(Color online) NRG conductance $G_\alpha(T)/G_0$ of the $S=1$ Kondo impurity model for several 
positive magnetic  anisotropies $D$ across the topological quantum phase transition. The black (red) 
curves correspond to $D < D_c$ ($D > D_c$). From top to bottom: $D=0.0,$ $1\times 10^{-4},$ 
$1.15\times 10^{-4},$ $1.3\times 10^{-4},$ $1.35\times 10^{-4},$ $1.3505\times 10^{-4},$ 
$1.351\times 10^{-4},$ $1.315\times 10^{-4},$ $1.352\times 10^{-4},$ $1.3525\times 10^{-4},$ 
$1.353\times 10^{-4},$ $1.3535\times 10^{-4},$ $1.354\times 10^{-4},$ $1.345\times 10^{-4},$ 
$1.355\times 10^{-4},$ $1.4\times 10^{-4},$ $1.6\times 10^{-4}.$ 
The parameters are $J=0.2$ and $W=1$. We take $\Lambda=3$ and keep 4000 NRG states. 
For these parameters, $T_K^0 = 4.2 \times 10^{-5}$ and $D_c = 1.352\times 10^{-4}$.}
\label{fig9:cond_dpos}
\end{figure}

To illustrate the generic appearance of this topological quantum phase transition for the $S=1$ impurity, 
we consider the Kondo limit of the two-orbital Anderson model (\ref{2orb-ham}), that is, a $S=1$ impurity 
coupled through an exchange interaction $J$ with two-degenerate conduction bands. 
In Fig.~\ref{fig9:cond_dpos}, we show the NRG conductance of 
the Kondo model as a function of temperature, for several positive $D$, below and above the critical 
anisotropy $D_c = 1.352\times 10^{-4}$ ($J=0.2$, $W=1$). 
It can be seen that for $D < D_c$, $G(T)$ reaches the unitary limit at low temperatures, corresponding to a fully-screened Kondo effect; on the other hand, for 
$D > D_c$, the conductance goes to zero, as the magnetic moment of the impurity is quenched by the single-ion anisotropy. Close to $D_c$, $G(T)$ exhibits an extended plateau at one-half of the unitary limit value, a typical characteristic of the 2CK effect. 

\begin{figure}[ht]
\includegraphics*[width=1.0\columnwidth]{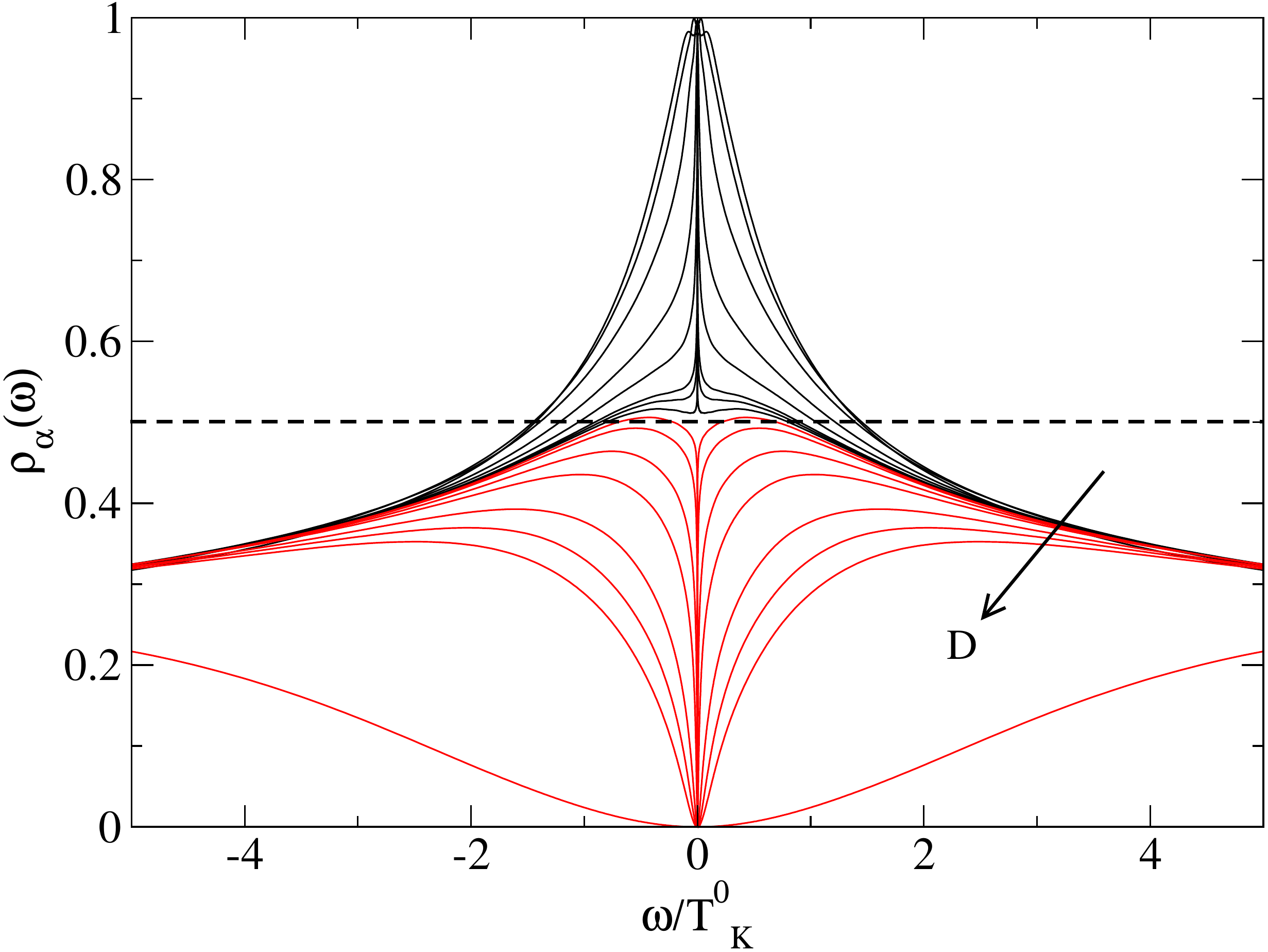}
\caption{(Color online) Normalized NRG spectral function $\rho_\alpha(\omega)$ of the $S=1$ Kondo impurity  model as a function of $T/T^0_K,$ for several positive $D$ across the topological quantum phase transition. 
The black (red) curves correspond to $D < D_c$ ($D > D_c$).
From top to bottom: $D= 0.0,$ $5\times 10^{-5},$ $8\times 10^{-5},$  
$1\times 10^{-4},$ $1.15\times 10^{-4},$ 
$1.2\times 10^{-4},$ $1.21\times 10^{-4},$  $1.22\times 10^{-4},$ 
$1.23\times 10^{-4},$ $1.25\times 10^{-4},$ 
$1.3\times 10^{-4},$ $1.35\times 10^{-4},$ 
$1.6\times 10^{-4},$ $1.8\times 10^{-4},$ $2\times 10^{-4},$ $5\times 10^{-4}.$ 
The parameters are $W=1, J=0.2$. We take $\Lambda=2$ and keep 3000 NRG states. For these parameters, $D_c= 1.225 \times 10^{-4}$ and $T_K^0 = 3.8\times 10^{-4}$.  }
\label{fig10:spectra_dpos}
\end{figure}

As another signature of the quantum phase transition that occurs for $D=D_c$, 
Fig.~\ref{fig10:spectra_dpos} displays the spectral density of states in the 
Kondo limit~\cite{mitchell10} calculated with NRG 
of the $S=1$ Kondo impurity model, around the Fermi level 
energy, for several positive $D$ across the critical one. For $D < D_c$ (black curves) the full screening of 
the $S=1$ impurity gives rise to a Kondo resonance, while for $D > D_c$ (red curves) a dip appears at the 
Fermi level due to the anisotropy-driven quenching of the magnetic degree of freedom of the impurity. 
At $D \simeq D_c$, the spectral density at $\omega=0$ takes half of its value in the fully screened Kondo 
phase. This is another indication of the emergence of 2CK physics at critical $D_c$. 

\begin{figure}[ht]
\includegraphics*[width=0.7\columnwidth]{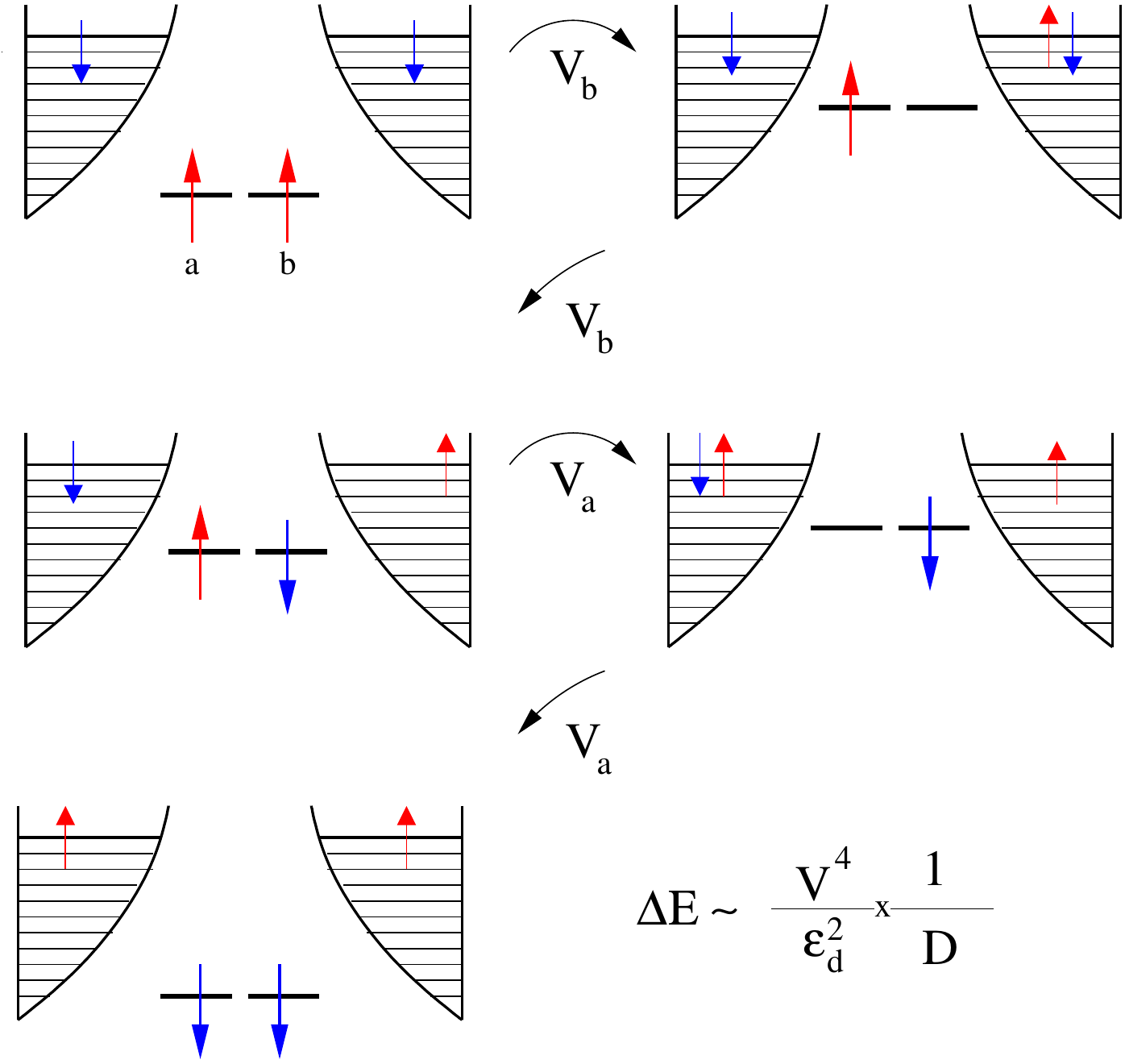}
\caption{(Color online) A fourth-order hybridization process that leads to an effective spin-flip between the $S_z=\pm 1$ projections for negative 
$D$.}
\label{fig9:4orden}
\end{figure}

With the aim of complementing the above mentioned results, now we analyze the behavior of 
our model~(\ref{2orb-ham}) with negative $D$. In this case, the isolated impurity has a doubly degenerate ground state, corresponding to $S_z = \pm 1.$
Although these spin projections differ in $|\Delta S_z| =2$ and, consequently, they cannot be connected by the usual second order hybridization processes 
that originate the Kondo exchange interaction, we have found that fourth-order hybridization processes (see Fig.~\ref{fig9:4orden}) lead to a fully Kondo 
screening of the $S=1$ impurity for negative $D$. 

\begin{figure}[ht]
\includegraphics*[width=0.8\columnwidth]{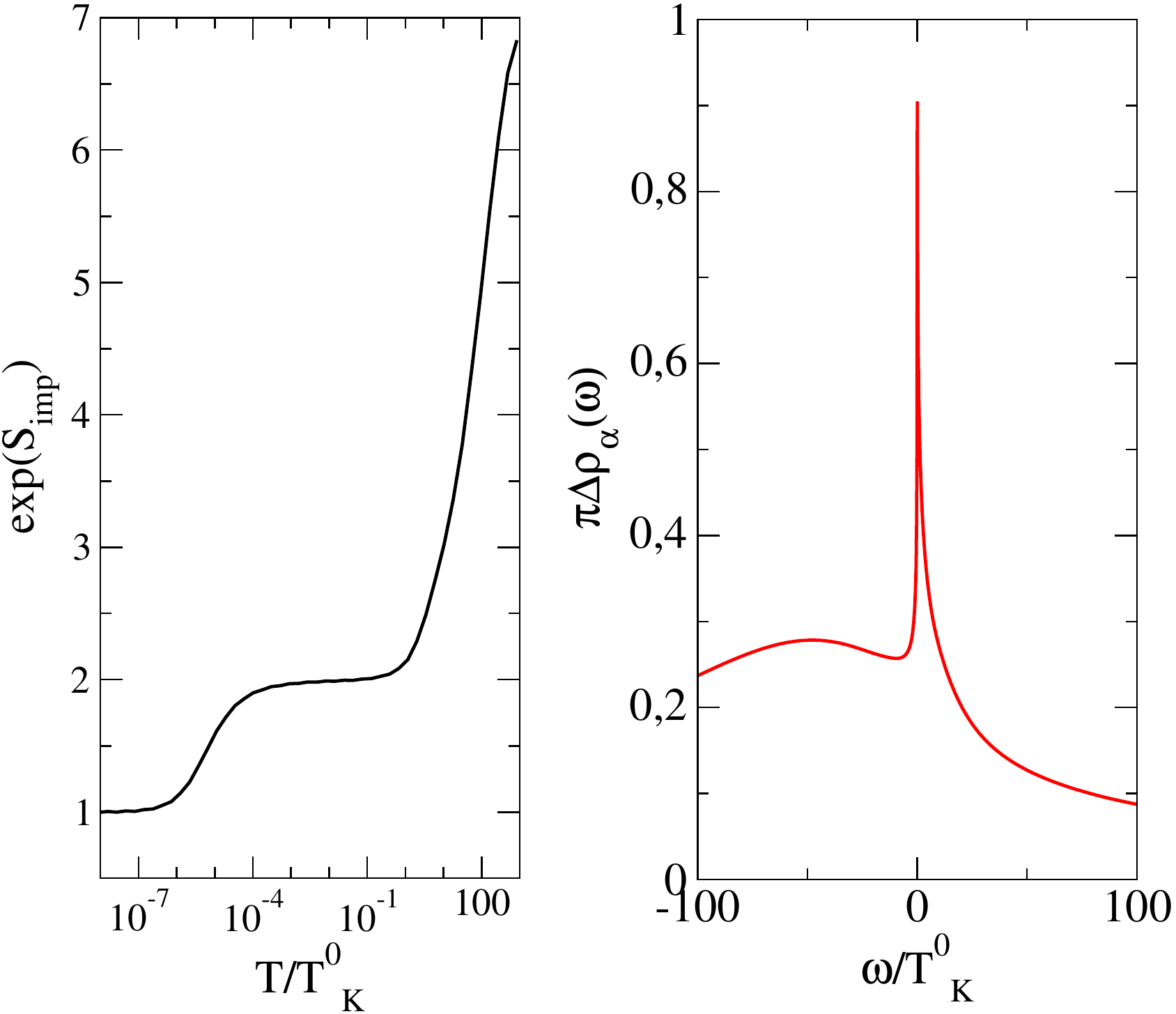}
\caption{(Left) NRG impurity entropy $S_{imp}$ as a function of $T/T^0_K$. (Right) NRG impurity spectral function $\rho_{\alpha\sigma}(\omega)$ 
as a function of $\omega/T^0_K$. The model parameters
are $\Delta=0.1$, $\epsilon_d=-0.2 \Delta$, with a corresponding $T^0_K \simeq 1.245\times 10^{-3}$, and  $D=-16 T^0_K$.}
\label{dneg_sr}
\end{figure}

The left panel of Fig.~(\ref{dneg_sr}) displays the NRG impurity entropy contribution as function of the normalized temperature $T/T^0_K$, 
for $D=-16T_K^0.$ It can be seen that, as for $D=0$,  at the higher temperatures $e^{S_{imp}}$ saturates at the value given by the localized Hilbert space.  
At intermediate temperatures, once the charge fluctuations are frozen, a plateau can be observed with $e^{S_{imp}} \simeq 2,$ corresponding to the double degenerate 
impurity $S_z = \pm 1$ degrees of freedom. As these magnetic states are fully Kondo screened at lower temperatures, 
the impurity entropy goes to zero. On the other hand, the right panel of Fig.~(\ref{dneg_sr}) shows the NRG impurity spectral function as function of frequency 
around the Fermi level, at a very low temperature. $\rho_{\alpha\sigma}$ has the typical structure of a Kondo state, and it can be seen that it
satisfies the Friedel sum rule. Both curves, entropy and spectral function, along the NRG spectra, indicates that for negative $D$ the ground state of 
model (\ref{2orb-ham}) is a conventional local Fermi liquid. 

\begin{figure}[ht]
\includegraphics*[width=0.80\columnwidth]{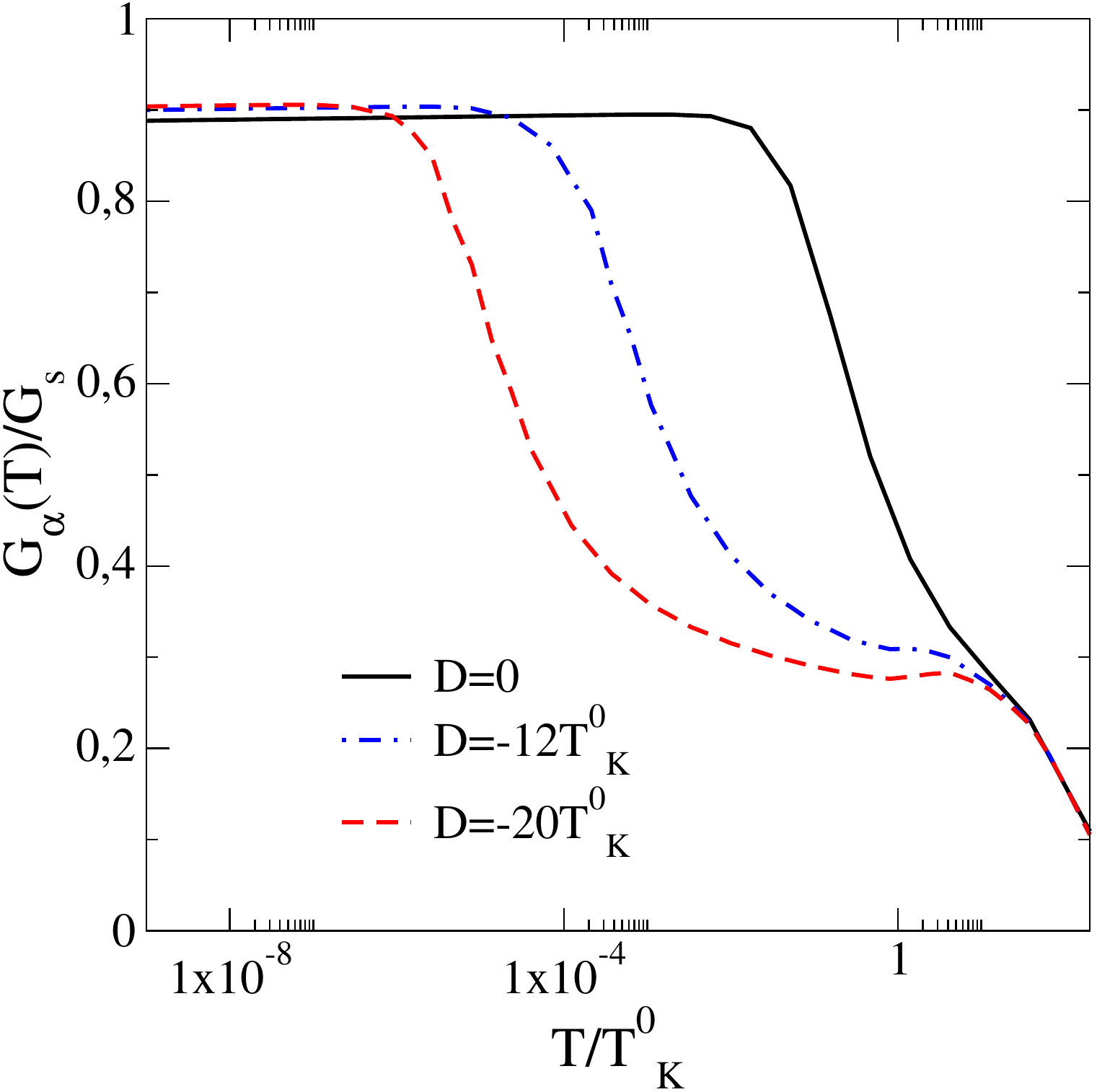}
\caption{(Color online) NRG conductance $G_\alpha(T)$ as a function of $T/T^0_K$ for three different negative magnetic 
 anisotropies $D$. The other parameters are the same as in Fig.~(\ref{dneg_sr}).}
\label{conductancia-anisotropia.eps}
\end{figure}

Finally, through the NRG conductance (see Fig.~(\ref{conductancia-anisotropia.eps}), we can estimate the Kondo temperature using the usual rule 
$G(T_K) = G(T\rightarrow 0)/2$. It can be seen that this energy scale $T_K(D)$ rapidly goes down as $D$ becomes more negative. In fact, 
$T_K(D)$ obeys an exponential scaling law with $|D|:$
\begin{equation}
 T_K(D) \propto T_K^0 {\rm exp}\left[-c\left(\frac{D}{T_K^0}\right)^2\right]
\end{equation}
with $c$ a constant of order of one.

\section{Conclusions} \label{sec_conclusions}
In this work, we have studied a spin-1 Anderson impurity model --in which the triplet ground state 
is mixed with a configuration with two doublets by means of two degenerate conduction channels--, that describes 
a single Ni impurity embedded in an O-doped Au chain~\cite{dinapoli15}. 
In agreement with the predictions of Nozi\`{e}res and Blandin~\cite{nozieres80}, this two-orbital two-channel impurity model exhibits 
a fully-screened Kondo effect at low temperatures. 
As the experimental realizations of the fully-screened Kondo effect for high-spin quantum dots are rare at present as 
compared to what happens in bulk systems, our results provide a useful guide to potential experimental studies 
of the Ni impurity in gold chains or related systems, allowing, for example, to discern the spin of the impurity. 
In this sense, to be more realistic, we have included the effect of a single ion anisotropy $D$ in the impurity.  
There is a plethora of experiments on nanoscopic systems, which cannot  be described with the single channel 
$S=1/2$ Anderson (or Kondo) model, in particular several experimental realizations with 
$S>1/2$~\cite{parks07,roch08,florens11,cornaglia11,roura09,roura10,expmn,orma} and also with degenerate 
orbitals~\cite{mina,joaq,moro} have been studied. Therefore, it is reasonable to expect that 
experimental realizations of fully compensated Kondo impurities will appear in the near future.

We have solved the impurity model using two methods that give very reliable information at complementary energy scales: 
the numerical renormalization group, numerically exact at low energies of the order of the Kondo temperature, 
and the non-crossing approximation, that takes correct account of the charge transfer processes at higher energies. 

For $D=0$, we have found the expected signatures of the local Fermi liquid behavior at low temperatures: 
the universality with a 
single Kondo energy scale, the vanishing of the impurity entropy contribution, the large electrical conductance in agreement 
with the Friedel sum rule, among others. However, the Kondo temperature is strongly reduced in comparison with the $s=1/2$ case, 
a phenomenon that has been experimentally observed in bulk systems since the 1960's~\cite{nevidomskyy09}. 
Furthermore, the empirical expression of the conductance as a function of temperature (Eq. \ref{ley_empirica}), that is used to 
experimentally discern the spin value of the impurity~\cite{parks10}, has a noticeable different fitting $s$ parameter as compared
with the fully screened spin-$1/2$ and the underscreened spin-$1$ Kondo cases. Another interesting feature is that the mixed 
valence regime seems to be much suppressed in the spin$-1$ case.

At higher energies, the charge transfer peak for the spin-$1$ model exhibits a very different behavior in comparison with the 
spin-$1/2$ model: the Haldane shift of the bare energy $\epsilon_d$ is cut in half and has an opposite sign 
($\epsilon_d^\ast$ is closer to the Fermi level), while the width of the charge transfer peak is reduced to $\simeq 3\Delta$. 
This last result, together with suppression of the mixed valence regime, points out that charge fluctuations are significantly reduced 
for high-spin impurities.

As found earlier~\cite{blesio18}, the single-ion anisotropy $D$ has a strong effect on the Kondo physics: while for any negative $D$ the Kondo 
effect survives,  with a reduced $T_K(D)$, for a critical positive $D_c$ there is a topological quantum phase transition, 
from the usual local Fermi liquid at lower $D$ to a topologically non-trivial non-Landau Fermi liquid for larger anisotropies. 
Just at the transition, the impurity shows the signatures of a non-Fermi liquid two-channel Kondo behavior. In this work
we have calculated the conductance and the spectral density in the Kondo limit very near the transition, 
showing the abrupt remarkable changes of both quantities at $D_c$.

For negative $D$ the Kondo effect always persists at low enough temperatures, but the Kondo temperature is 
strongly reduced because the remaining degenerate states of the impurity are mixed by a higher (fourth)
order process in the hybridization between impurity and conduction electrons.

We hope that our detailed study of the spin$-1$ Anderson impurity model encourages the experimental search of low dimensional high-spin 
fully-screened Kondo systems, like the proposed Ni impurity in O-doped Au-chain. As we have shown, there are several observables 
that can be used to differentiate the high-spin and the usual spin$-1/2$ cases.

\acknowledgments
This work was partially supported by PIP 112-201501-00506 of CONICET (Argentina), and PICT 2013-1045 
of the ANPCyT (Argentina).

\appendix

\section{Kondo temperature within the NCA approximation}\label{app_kondo_nca}
In this Appendix, we show that the Kondo scale is overestimated within the non-crossing approximation.

In the isotropic case, $D=0$, and for degenerate levels (as for the Ni impurity in the O-doped Au chain), 
the system of equations that determines the NCA self-energies is reduced to
\begin{eqnarray} \label{nca-selfenergies-simetrico}
\Sigma_{1}(\omega)&=&\frac{3\Delta}{2\pi}\int_{-W}^{W}~d\epsilon f(\epsilon) G_{2}(\epsilon+\omega),\nonumber \\
\Sigma_{2}(\omega)&=&\frac{2\Delta}{\pi}\int_{-W}^{W}~d\epsilon f(\epsilon) G_{1}(\omega+\epsilon),
\end{eqnarray}
where $G_{2}(\omega)$ and $G_{1}(\omega)$ represent the Green's function of the triplet and both doublets 
$\alpha = xz, yz$ components, respectively. 

With the help of the redefinition $\Delta=2\Delta'$ for the hybridization, it can be seen that the set of 
self-energies equations is the same as for the NCA treatment of the $SU(N) \times SU(M)$ generalization of 
the multichannel single-impurity Kondo model, with  $M=4$ identical conduction bands and being $N=3$ the degeneracy 
in the impurity spin quantum number~\cite{cox93}, 
\begin{eqnarray} \label{nca-selfenergies-simetrico2}
\Sigma_{1}(\omega)&=&\frac{3\Delta'}{\pi}\int_{-W}^{W}~d\epsilon f(\epsilon) G_{2}(\epsilon+\omega),\nonumber \\
\Sigma_{2}(\omega)&=&\frac{4\Delta'}{\pi}\int_{-W}^{W}~d\epsilon f(\epsilon) G_{1}(\omega+\epsilon).
\end{eqnarray}

For this system, the characteristic energy scale $T^{\rm NCA}_{K}$ can be obtained analytically from 
the zero temperature limit of the self-energies, and it is given by the expression~\cite{kim97,cox93}
\begin{equation}
T^{\rm NCA}_{K} = W \left(\frac{\Delta'}{\pi W}\right)^{\frac{M}{N}} \times {\rm exp}\left(\frac{\pi\epsilon_d}{N\Delta'}\right). 
\end{equation}
For the simplest case of the one-channel ($M=1$), infinite Coulomb repulsion, and spin $s=1/2$ ($N=2$) 
Anderson model, NCA gives the Kondo temperature
\begin{equation}
 T_{K,\rm NCA}^{s=1/2} = \sqrt{\frac{\Delta W}{\pi}}e^{\pi\epsilon_d/2\Delta}. 
\end{equation}
This value coincides with the exact Kondo energy except for the prefactor $1/\sqrt{\pi}\approx 0.5$.

On the other hand, for the case of interest we derive the following NCA Kondo scale,
\begin{equation}
 T_{K, NCA}^{S=1} =W \left(\frac{\Delta}{2\pi W}\right)^{4/3}~e^{2\pi\epsilon_d/3\Delta}, 
\end{equation}
which can be written in terms of $T_{K}^{s=1/2}$ and in units of $W$ as follows
\begin{eqnarray}
 T_{K, NCA}^{S=1} = \sqrt{8}\left(\frac{\Delta}{2\pi}\right)^{2/3}~\left(T_{K}^{s=1/2}\right)^{4/3}.
\end{eqnarray}

The relation with the asymptotic Kondo temperature for the spin $S=1$ obtained from the NRG calculations
in Eq.~(\ref{kondo-scale-spin-1}), $c\sqrt{\Delta}~\left(T_{K}^{s=1/2}\right)^2$, shows that the NCA largely 
overestimates the Kondo scale as
\begin{eqnarray}
 \frac{T_{K}^{S=1}}{T_{K, NCA}^{S=1}} \approx c (2\pi)^{4/3} \Delta^{1/6} e^{\pi \epsilon_d/3\Delta} \ll 1, 
\end{eqnarray}
for typical $\epsilon_d$ and $\Delta$ values in the Kondo regime.
\begin{center}
 \textit{Intensity of the Kondo peak within the NCA approximation}\\
\end{center}

For such a model, the NCA spectral density at the Fermi level is expected to be 
$\rho(0)\sim\frac{2\pi}{(N+M)^2 \Delta'}$, (see Appendix B of Ref.~\onlinecite{kim97}). 
We have verified that our calculations satisfy this rule (An additional factor 2 was 
included due to the definition of the physical operator).

\end{document}